\documentstyle[aps,epsfig]{revtex}

\begin{document}

\title{OCS in small para-hydrogen clusters:
energetics and structure with $N$=1-8 complexed hydrogen molecules}

\author{F. Paesani, R.~E.~Zillich and K.~B.~Whaley}

\address{Department of Chemistry and 
Pitzer Center for Theoretical Chemistry, University of California,
Berkeley, CA 94720}
\maketitle

\begin{abstract}
We determine the structure and energetics of complexes of the 
linear OCS molecule with small numbers of para-hydrogen molecules, 
$N$=1-8, using zero temperature quantum Monte Carlo methods.  
Ground state calculations are carried out with importance-sampled rigid 
body diffusion Monte Carlo (IS-RBDMC) and excited state calculations 
with the projection operator imaginary time spectral evolution (POITSE) 
methodology.  The ground states are found to be highly structured, 
with a gradual build up of two axial rings as $N$ increases to 8.  
Analysis of the azimuthal density correlations around the OCS molecule 
shows that these rings are quite delocalized for small $N$ values, 
but become strongly localized for $N\geq 5$ .  
Excited state calculations are made for a range of total cluster 
angular momentum values and the rotational energy levels fitted to obtain 
effective rotational and distortion constants of the complexed OCS 
molecule as a function of cluster size $N$.  Detailed analysis of these 
spectroscopic constants indicates that the complexes of OCS with 
para-hydrogen have an unusually rich variation in dynamical behavior, 
with sizes $N$=1-2 showing near rigid behavior, sizes $N$=3-4 
showing extremely floppy behavior, and the larger sizes $N$=5-8 
showing more rigid behavior again.
The large values of the distortion constant $D$
obtained for $N$=3-4 are rationalized in terms of the coupling
between the OCS rotations and the "breathing" mode of the first,
partially filled ring
of para-hydrogen molecules.

\end{abstract}

\section{Introduction}

Complexes of small molecules with variable number of para-hydrogen ($p$-H$_2$) molecules 
pose a challenging and fascinating arena for the manifestation of nuclear 
quantum effects on structure and spectroscopy.  Like their better known analogous 
helium clusters, para-hydrogen clusters are expected to be strongly influenced 
by the quantum nature of the $p$-H$_2$ molecules.  The mass of $p$-H$_2$ being one half 
that of $^4$He actually suggests that larger zero point effects might be expected.  
However, the greater binding of $p$-H$_2$ to itself and to other species  
means that the greater delocalization tendency of $p$-H$_2$ 
competes with greater localizing potential energy terms.  The competition between 
these two strong effects is what causes both calculation and analysis of 
para-hydrogen clusters and complexes to be considerably more difficult 
than that for analogous helium systems.  Like helium, para-hydrogen is a boson, 
but because of the greater binding energy, $p$-H$_2$ solidifies in the bulk 
before cooling to a low enough temperature for macroscopic manifestation 
of the boson permutation exchange symmetry to occur with a superfluid phase.  
However, in finite clusters vestiges of superfluidity may occur, particularly 
when the para-hydrogen packing is expanded or otherwise constrained by the reduced 
dimensionality and/or the $p$-H$_2$ molecules are bound to foreign species.  

The linear OCS molecule has played a key role in developing our understanding 
of the fundamental forces controlling the properties of $p$-H$_2$ complexes. 
Experimental studies with OCS complexed by variable numbers of para-hydrogen 
($N \leq 16$) and embedded in helium droplets have shown a very rich spectroscopic behavior, 
with both vibrational shifts and rotational fine structure providing indirect information 
on the symmetry and distribution of the $p$-H$_2$
around the OCS molecule \cite{grebenev00,grebenev01,grebenev02}. 
In helium, the rotational spectra of the para-hydrogen complexes are well fit 
by asymmetric tops for $N$=1-4, while larger clusters appear adequately described 
by a symmetric top Hamiltonian \cite{grebenev01}. 
For certain sizes, $N$=5, 6, 11, 14-16, no $Q$-branch is seen in the spectra.  
For the larger sizes, $N$=14-16, this disappearance is temperature dependent, 
which led to the suggestion that it might be a manifestation of superfluidity 
in the para-hydrogen component.  Path integral Monte Carlo calculations 
for a complete solvation shell ($N$=17) have shown that a transition to an anisotropic 
superfluid state is indeed found at temperatures below T$=0.3$~K, and that this 
can account for the disappearance of the $Q$-branch for $N \geq 11$ \cite{kwon02}.  
However, for the smaller sizes, the spectral anomaly is independent of temperature.  
In the absence of detailed knowledge of the structure and rigidity of the complex, 
the microscopic origin of this anomaly is less clear, 
although the permutation symmetry can also be expected to play a role.

Much less is known experimentally about the complexes of OCS with para-hydrogen 
in the absence of a solvating helium environment.  The OCS($p$-H$_2$) dimes has been 
characterized by high resolution spectroscopy \cite{tang02}, and its strcuture has been shown 
to be approximate with that of an asymmetric top rotor.  The structure of this dimer is very similar 
in the gas phase and in a helium droplet \cite{grebenev01a,kwon03}. 
For complexes with larger numbers of para-hydrogen molecules, 
no spectroscopic data have been published. 
We expect that OCS complexed by 1~$<N<$~17 para-hydrogen molecules will show 
a very rich variation in structure, energetics, and spectroscopic properties, 
as a consequence of the strongly modulated solvating hydrogen density along 
the axis of the OCS molecule seen in path integral calculations for $N=$17 \cite{kwon02}.
These modulations are considerably stronger than those found in the analogous 
complexes of OCS with helium \cite{paesani01,kwon01,paesani03} 
because of the stronger ($p$-H$_2$)-OCS and ($p$-H$_2$)-($p$-H$_2$) interactions.  
Consequently we can expect a considerably more complex size dependence 
of the spectroscopic properties for the complexes with $p$-H$_2$ as the first 
solvation shell is filled.  The small complexes of OCS with helium have recently 
been shown to undergo a transition from near rigid molecular complex behavior 
to true quantum (or "superfluid") solvation characterized by permutation 
exchanges along the molecular axis, as the number of helium atoms increases 
from $N$=1 to $N$=20 \cite{paesani03}. 
Given the recent demonstration of anisotropic superfluidity in OCS($p$-H$_2$)$_{17}$ \cite{kwon02}, 
we expect an analogous transition for complexes with molecular hydrogen.  
However, the considerably greater competition between quantum (kinetic) delocalization 
and (potential) localization for para-hydrogen renders even qualitative prediction 
of the spectroscopic behavior impossible without detailed microscopic calculations.  
Furthermore, for the intermediate sizes 1~$<N<$~17, the properties of gas phase 
OCS($p$-H$_2$)$_{17}$ complexes and their analogous embedded in helium droplets 
can be expected to show maximal differences, since this is the size regime 
in which subtle changes in para-hydrogen and helium density distributions 
can give rise to significant energy differences.  

In this paper we present results of ground and excited state calculations 
for OCS complexed with $N$=1-8 para-hydrogen molecules, as part of a 
larger systematic study of the complexes with up to a complete solvation shell \cite{paesani03a}. 
We employ two kinds of zero temperature quantum Monte Carlo methods, 
the diffusion Monte Carlo (DMC) and the projection operator imaginary 
time spectral evolution (POITSE) approach.  
These microscopic calculations provide energy levels and 
rotational and distortion constants, in addition to energetics 
and ground state structures.  
We find a very interesting variation in structure for this series of small complexes, 
showing considerably more inhomogeneous behavior than the analogous series 
of complexes of OCS with $^4$He.  We determine the structure both along the molecular 
axis and around it, with analysis of para-hydrogen pair correlation functions 
and find that this allows us to provide a good description of the extent 
of localization in the ground state.  
The excited state energies are extracted from a maximum entropy analysis 
of the imaginary time correlation functions derived from the POITSE approach.  
We find that a simple multi-exponential fit of these 
correlations functions provides similar 
results, but with larger uncertainties.   
Analysis of the fitted spectral constants resulting from the excited state 
energies reveals a gradual change of symmetry of the OCS($p$-H$_2$)$_{N}$ 
complexes as $N$ increases, 
which is accompanied by a monotonic decrease in the effective rotational 
constant over this size range.  
While overall the analysis of spectral constants indicates
more rigid structures for these OCS complexes with para-hydrogen than
those with helium, the fitted spectral constants nevertheless show
a markedly strong appearance of floppy, non-rigid structures 
for both the $N$=3  and $N$=4 complexes.  
We discuss these structures and predicted spectroscopic constants in 
relation to available experimental data and in
the context of the analogous OCS($^4$He)$_N$ complexes \cite{paesani03}.
  
\section{Theoretical approach}
\subsection{Interaction Potentials}
The total interaction potential for the OCS($p$-H$_2$)$_N$ clusters is obtained as a 
sum of all pairwise contributions, neglecting the effects of three-body
forces among the para-hydrogen molecules, while correctly treating
the many-body interaction between the linear OCS impurity molecule 
and each $p$-H$_2$ 
molecule one at a time.
Because of its nuclear spin symmetry (I=0), para-hydrogen can only access 
even values of its angular momentum, i.e., $j_H$=0, 2, 4, etc.
Furthermore, given the large spacing between the rotational
levels, $j_H$ can generally be assumed to be a good quantum number 
in the weakly bound van der Waals 
complexes ~\cite{mckellar90,mckellar98,tang02}. These considerations, together with the fact that our 
calculations are carried out at a temperature of 0~K, 
allow us to consider each $p$-H$_2$ molecule as being in its 
$j_H$=0 rotational state. Therefore, in the present study, we treat the
para-hydrogen molecules as spherical particles.

\begin{figure}[ht]
\epsffile{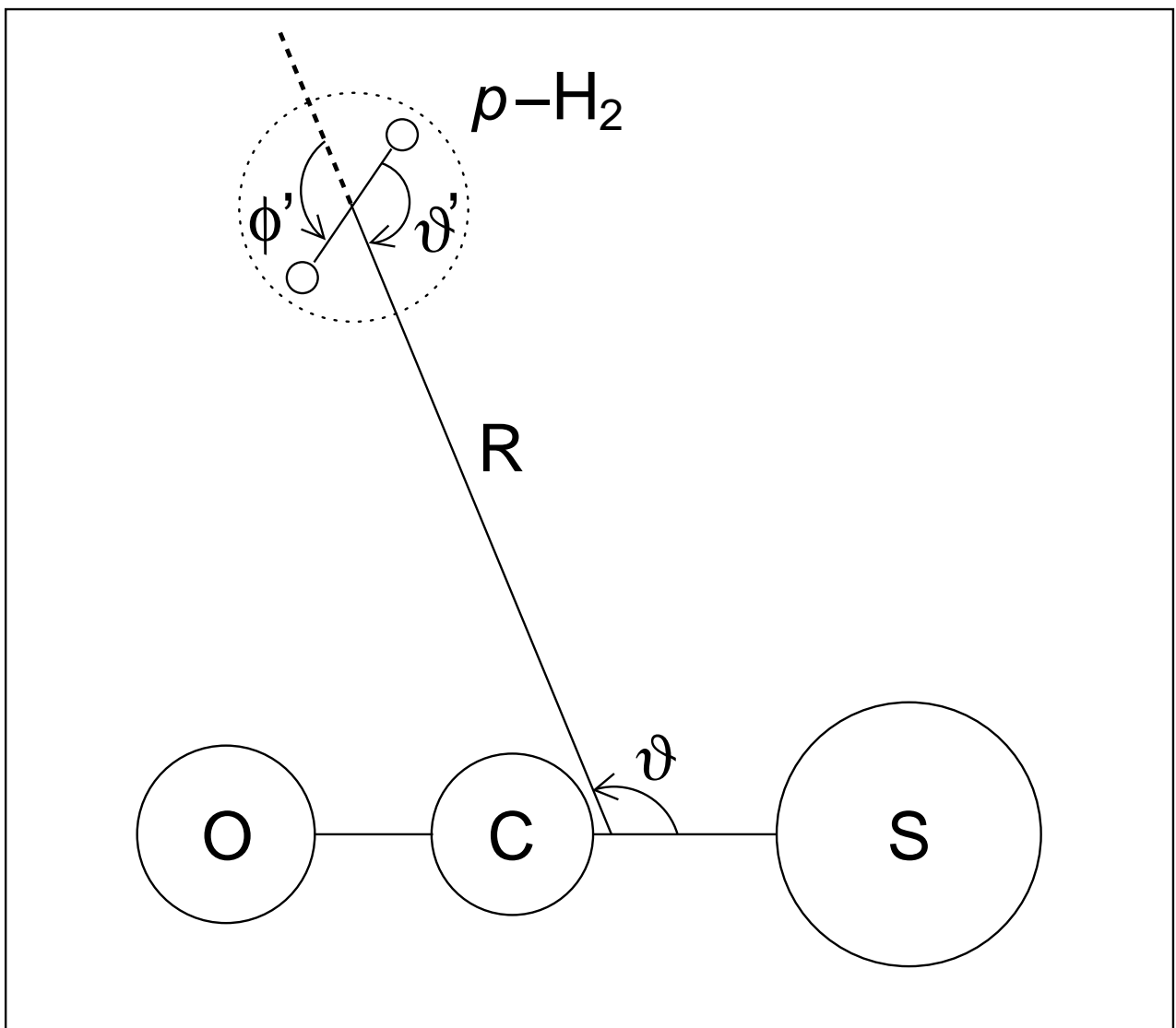} 
\caption{Jacobi coordinates for the OCS-($p$H$_2$) complex.
\label{fig:coordinates}
}
\end{figure}

The total interaction potential can then be expressed as
\begin{equation}
V({\bf R})=\sum_{i=1}^N V^{OCS-(p-H_2)}(R_i,\vartheta_i)
          +\sum_{i<j} V^{(p-H_2)-(p-H_2)}(R_{ij})
\label{eq:potential}
\end{equation}
where ${\bf R}$ is a generic vector that defines the coordinates of all the particles. Therefore,
$R_{ij}$ is the distance between $p$-H$_2$ molecules $i$ and $j$, and $R_i$,
$\vartheta_i$ are the Jacobi coordinates of the $i$th para-hydrogen molecule in the
center of mass frame of the OCS (see Fig.~\ref{fig:coordinates}).

For the ($p$-H$_2$)-($p$-H$_2$) interaction we use the spherical part of the
empirical potential proposed by Buck {\em et al.} \cite{buck83}.
The OCS-($p$-H$_2$) interaction is obtained after integration over the $p$-H$_2$ angular
variables ($\vartheta',\phi'$) of a 4-dimensional {\em ab initio} 
potential energy surface recently calculated
by Higgins {\em et al.} using fourth-order 
M\"oller-Plesset perturbation theory \cite{higgins}. A contour
plot of the averaged potential is shown in Fig.~2a. The global
minimum of -144.5 cm$^{-1}$ is located at $R=3.35$~\AA~and $\vartheta=105^\circ$. 
Two other local minima
corresponding to the two collinear geometries are also found. 
The first, at $\vartheta=0^\circ$ and
$R=4.52$~\AA~has a well depth of -91.4 cm$^{-1}$. The second, with a well depth of -69.2 cm$^{-1}$ 
is located at $\vartheta=180^\circ$ and $R=4.92$~\AA. In Fig.~2b we show the minimum
potential energy angular path, from which we see that a 
saddle point of -66.49 cm$^{-1}$ is located at $\vartheta=53.5^\circ$ and $R=4.29$~\AA.

\subsection{Ground state calculations}
\subsubsection{Variational Monte Carlo (VMC) method}
The ground state energies for the OCS($p$-H$_2$)$_N$ clusters are first calculated
using the VMC method. In this stochastic approach,
one constructs a trial wavefunction $\Psi_T({\bf R};\{\bf{p}\})$ which approximates
the exact ground state wavefunction of the system, $\Phi_0({\bf R})$,
for a given Hamiltonian $\hat{H}$.
Here, $\{{\bf p}\}$ denotes 
a set of adjustable parameters that control the shape of the trial wavefunction.
The energy expectation value is then evaluated using a Monte Carlo integration.
In most cases, $\Psi_T({\bf R};\{{\bf p}\})\neq\Phi_0({\bf R})$, and, consequently,  
the VMC method provides an upper bound for the total energy,
\begin{equation}
E_0 \leq E(\{{\bf p}\}) = 
              {{\int \Psi^*_T({\bf R};\{{\bf p}\}) \hat{H} \Psi_T({\bf R};\{{\bf p}\}) d{\bf R}}
               \over
               {\int \mid \Psi_T({\bf R};\{{\bf p}\}) \mid^2 d{\bf R}}}.
\label{eq:E_av}
\end{equation}
\noindent
The last equation can be rewritten in a convenient form for Monte Carlo
integration as
\begin{eqnarray}
E(\{{\bf p}\}) &=& {{\int \mid \Psi_T({\bf R};\{{\bf p}\}) \mid^2
                     \Psi^{-1}_T({\bf R};\{{\bf p}\}) \hat{H} \Psi_T({\bf R};\{{\bf p}\}) d{\bf R}}
                   \over
                   {\int \mid \Psi_T({\bf R};\{{\bf p}\}) \mid^2 d{\bf R}}}  \\
               &\simeq& {{1}\over{n}} \sum_{k=1}^{n} E_{L}({\bf R}_k;\{{\bf p}\})
\label{eq:E_vmc}
\end{eqnarray}
where $n$ is the total number of configurations that are sampled
from the probability density function 
$\mid \Psi_T({\bf R};\{{\bf p}\}) \mid^2$ using the Metropolis algorithm \cite{metropolis53}. 
In eq.~(\ref{eq:E_vmc}) $E_{L}({\bf R}_k;\{{\bf p}\})$ is the local energy
\begin{equation}
E_{L}({\bf R}_k;\{{\bf p}\}) = \Psi^{-1}_T({\bf R};\{{\bf p}\}) \hat{H} \Psi_T({\bf R};\{{\bf p}\}).
\label{eq:E_local}
\end{equation}
\noindent
In our implementation $\Psi_T({\bf R};\{{\bf p}\})$ is given by the usual generalized product
form
\begin{equation}
\Psi_T({\bf R};\{{\bf p}\})= \prod_{i<j}^{N} \xi_T(R_{ij}; {\bf p'})
                             \prod_{i=1}^{N} \chi_T(R_i,\vartheta_i; {\bf p})
\label{eq:prod_func}
\end{equation}
where $R_i$, $\vartheta_i$ and $R_{ij}$ are the same as in eq.~(\ref{eq:potential}).
This type of trial function is symmetric with respect to $p$-H$_2$ molecule exchanges and
thus explicitly includes Bose statistics.

For the ($p$-H$_2$)-($p$-H$_2$) correlation, we used the two-parameter function
\begin{equation}
\xi_T(R_{ij}) = \exp{(-\frac{p'_1}{R_{ij}^5}-p'_2R_{ij})}.
\label{eq:h2_func}
\end{equation}
For the ($p$-H$_2$)-OCS correlation, 
we started from an analytic fit to the ground state wavefunction of the $N$=1
($p$-H$_2$)-OCS complex calculated by use of the collocation method \cite{cohen90}. 
This function has the form
\begin{eqnarray}
\chi_T(R_i,\vartheta_i) =
 \exp{\{ p_1r^{p_2} + p_3[p_9+p_4(\cos{\gamma}-p_8)^2]\ln{r}
       + [p_5r^2(\cos{\gamma}-p_8)^2-p_{10}]e^{p_6-p_7r}\}}
\label{eq:ocs_func}
\end{eqnarray}
where $r=\sqrt{y^2+z^2}$, $y=R_i\sin{\vartheta_i}-p_{12}$, $z=R_i\cos{\vartheta_i}-p_{11}$ and
$\cos{\gamma}=z/r$.

Optimization of the parameters $\{{\bf p}\}$ for $N>$1 is then performed by simultaneous minimization
of the energy $E(\{{\bf p}\})$ and its variance. This approach provides the best wavefunction
$\Psi_T({\bf R};\{{\bf p}\})$ given a particular functional form \cite{barnett93b}. 
The values of the optimized parameters are reported in Table~\ref{tab:parameters}.

In the present study, a VMC run typically consists of 5$\times$10$^6$ steps 
whose size is uniformly distributed and  
chosen such that the number of accepted moves is approximately mantained at half the number
of the attempted ones.

\subsubsection{Diffusion Monte Carlo (DMC) method}
Exact calculations of the ground state properties for a many-body quantum system
can be made by use of the DMC method \cite{hammond94}.
We employ here the 
rigid-body diffusion Monte Carlo (RB-DMC) scheme that is described 
in ref.~\cite{viel02}. In the following we provide only a summary of the main concepts
and give details specific to the OCS($p$-H$_2$)$_N$ system.

Within the RB-DMC formulation we treat the OCS molecule as a rigid-rotor interacting with 
$N$ spherical $p$-H$_2$ molecules.  
The imaginary time ($\tau=it/\hbar$)
Schr\"odinger equation can then be written as:
\begin{equation}
{\partial{\Phi({\bf R}, \tau)}\over\partial{\tau}}=
               -[\hat{H}-E_{ref}]\Phi({\bf R}, \tau)
\label{eq:it_schr}
\end{equation}
where $E_{ref}$ is the reference energy defining the zero of the absolute energy scale,
and $\hat{H}$ is the Hamiltonian of the system 
\begin{equation}
\hat{H}=-{\hbar^2\over 2M_0}\nabla_0^2-B_0\Big({\partial^2\over\partial\varphi_x^2}
      +{\partial^2\over\partial\varphi_y^2}\Big)
      -{\hbar^2\over 2m}\sum_{i=1}^N\nabla_i^2+V({\bf R}).
\label{eq:hamilton}
\end{equation}
In the r.h.s. of eq.~(\ref{eq:hamilton}) the first two terms
are the translational and the rotational kinetic energy of the OCS molecule, respectively.
$M_0$ is the mass of the OCS and $B_0$ its rotational constant. 
$\phi_x$, $\phi_y$ describe the OCS rotations
about the $x$- and $y$-axis in the molecule-fixed frame, where the $z$-axis is taken to lie
along the OCS axis.
The term $-{\hbar^2\over 2m}\nabla_i^2$ 
is the kinetic energy of the $i$th $p$-H$_2$ molecule and $m$ is the $p$-H$_2$ mass. 
Finally, $V({\bf R})$ is the total interaction potential of eq.~(\ref{eq:potential}).

It is well known that
efficiency of the DMC method can be improved if one
introduces a guiding function 
$\Psi_T({\bf R})$. 
For our OCS($p$-H$_2$)$_N$ system this
leads to the importance-sampled RB-DMC equation for the product function 
$f({\bf R},\tau)=\Phi({\bf R}, \tau)\Psi_T({\bf R})$, which differs from
eq.~(\ref{eq:it_schr}) by the presence of additional drift terms:
\begin{eqnarray}
{{\partial{f({\bf R}, \tau)}}\over{\partial\tau}} =&&
    {\hbar^2\over 2M_0}\nabla_0^2 f({\bf R}, \tau)
   -{\hbar^2\over 2M_0}\nabla_0 [f({\bf R}, \tau) {\bf F}_0^{(t)}({\bf R})] \nonumber \\
&& +B_0\Big({\partial^2\over\partial\varphi_x^2}+{\partial^2\over\partial\varphi_y^2}\Big)f({\bf R}, \tau)
   -B_0\Big({\partial\over\partial\varphi_x}+{\partial\over\partial\varphi_y}\Big)[f({\bf R}, \tau) {\bf F}_0^{(r)}({\bf R})]
 \nonumber \\
&& +{\hbar^2\over 2m}\sum_{i=1}^N\nabla_i^2 f({\bf R}, \tau)
   -{\hbar^2\over 2m}\sum_{i=1}^N\nabla_i [f({\bf R}, \tau) {\bf F}_i({\bf R})] \nonumber \\
&& -[E_L({\bf R})-E_{ref}] f({\bf R}, \tau).
\label{eq:is_schr}
\end{eqnarray}
Here, $E_L({\bf R})$ is the local energy of eq.~(\ref{eq:E_local}), and
\begin{eqnarray}
{\bf F}_0^{(t)}({\bf R}) & = & \nabla_0 \ln \mid \Psi_T({\bf R}) \mid^2 \\
{\bf F}_0^{(r)}({\bf R}) & = & \Big({\partial\over\partial\varphi_x}+{\partial\over\partial\varphi_y}\Big)
                                \ln \mid \Psi_T({\bf R}) \mid^2 \\
{\bf F}_i({\bf R}) & = & \nabla_i \ln \mid \Psi_T({\bf R}) \mid^2 .
\end{eqnarray}
The last three equations define the quantum forces that direct 
the sampling of $f({\bf R},\tau)$ into
regions where $\Psi_T({\bf R})$ is large.
               
The steady state solution $f({\bf R},\tau\rightarrow\infty)=\Phi_0({\bf R})\Psi_T({\bf R})$ of 
eq.~(\ref{eq:is_schr}) is obtained by
use of a random walk technique. 
An ensemble of walkers $\{{\bf X}_k\}$ with associated statistical weights $\{w_k\}$
is propagated in imaginary time from an
initial arbitrary distribution using the short time approximation of the Green's function
appropriate to eq.~(\ref{eq:is_schr}) \cite{viel02}. Here, ${\bf X}$ is a vector
in a (3$N$+5)-dimensional space. Two of these dimensions describe the OCS rotational motion
and the remainder 3($N$+1) the translational motion of all the particles. 
The propagation from $\tau$ to $\tau+\Delta\tau$ is achieved by a combination of
random gaussian displacements and systematic moves under the influence of the quantum forces.
Detailed balance is ensured by use of a generalized Metropolis scheme at each time step 
\cite{reynolds82}.

The ground state energy $E_0$ can then be calculated from averaging the local energy 
over the asymptotic distribution,
according to
\begin{eqnarray}
\langle E_L \rangle &=&{{\int f({\bf R},\tau\rightarrow\infty)E_L({\bf R}) d{\bf R}}
                        \over{\int f({\bf R},\tau\rightarrow\infty)d{\bf R}}}  \\
                    &=&{{\int \Phi_0({\bf R})\hat{H}\Psi_T({\bf R}) d{\bf R}}
                        \over{\int \Phi_0({\bf R})\Psi_T({\bf R}) d{\bf R}}}   \\
                    &=&E_0,
\end{eqnarray}
where the last equality follows from acting with $\hat{H}$ to the left.

Expectation values of coordinate operators $\hat{A}\equiv A({\bf R})$ can also be obtained by averaging
over $f({\bf R}, \tau\rightarrow\infty)$, but the integration leads here to a "mixed" expectation value
\begin{equation}
\langle\hat{A}\rangle_{mix}={{\int A({\bf R}) f({\bf R},\tau\rightarrow\infty}) d{\bf{R}}\over
                            {\int f({\bf R},\tau\rightarrow\infty}) d{\bf R}}
                           ={{\int \Phi_0({\bf R})A({\bf R})\Psi_T({\bf R}) d{\bf R}}\over
                            {\int \Phi_0({\bf R})\Psi_T({\bf R}) d{\bf R}}}.
\label{eq:mix}
\end{equation}
The "pure" expectation values can be retrieved after insertion 
of the factor $[\Phi_0({\bf R})/\Psi_T({\bf R})]$
in eq.~(\ref{eq:mix}), i.e.,
\begin{equation}
\langle\hat{A}\rangle_{pure}={{\int [\Phi_0({\bf R})/\Psi_T({\bf R})]A({\bf R}) 
           f({\bf R},\tau\rightarrow\infty}) d{\bf{R}}\over
           {\int [\Phi_0({\bf R})/\Psi_T({\bf R})]f({\bf R},\tau\rightarrow\infty}) d{\bf R}}
          ={{\int \Phi_0({\bf R})A({\bf R})\Phi_0({\bf R}) d{\bf R}}\over
           {\int \Phi_0({\bf R})\Phi_0({\bf R}) d{\bf R}}}
\label{eq:pure}
\end{equation}
In the present study the ratio $[\Phi_0({\bf R})/\Psi_T({\bf R})]$ is 
evaluated by use of the descendant weighting 
procedure as described in ref.~\cite{casulleras95}. All distributions 
shown here are computed with this method.
We shall analyze specifically the two dimensional density 
distribution $\rho(r,z)$ of the $p$-H$_2$ molecules
around the OCS molecule,
\begin{equation}
\rho(r,z)=\sum_i^N\Big\langle{{\delta(r_i-r)}\over{2\pi r}}\delta(z_i-z)\Big\rangle,
\label{eq:2D_rho}
\end{equation} 
where $z=R\cos\vartheta$ is defined to lie along the OCS axis, 
and the polar radius $r=R\sin\vartheta$ is perpendicular to this axis.
We shall also show the pair angular distribution function:
\begin{equation}
P_{12}(\phi)={{1}\over{N(N-1)}}\sum_{i<j}^N\langle\delta(\phi_{ij}-\phi)\rangle,
\label{eq:pair}
\end{equation}
normalized such that
\begin{equation}
\int P_{12}(\phi) d\phi=1.
\label{eq:pair_norm}
\end{equation}
Here $\phi_{ij}$ is the angle between two $p$-H$_2$ molecules and the OCS axis,
in a plane perpendicular to the molecular axis. 
Therefore, $P_{12}(\phi)$ is a three body distribution function integrated over all
variables with the constraint of fixed $\phi$. It describes the probability
to find a $p$-H$_2$ molecule at an angle $\phi$ from another molecule that is
arbitrarily located 
at $\phi$=0$^\circ$.  

In all the DMC calculations presented here, the guiding function $\Psi_T({\bf R})$ is
given by eqs.~(\ref{eq:prod_func}-\ref{eq:ocs_func}). We use
an ensemble of 1000 walkers whose initial distribution is taken from a VMC run.
A variable weight $w_k(\tau)$, with $w_k(0)=1$, is assigned to each walker. 
During the imaginary time propagation the weights can vary between $w_{min}$ and
$w_{max}$ according to the mixed weights/branching scheme described in 
ref.~\cite{barnett93b}, and the reference energy is updated continously according
to the growth energy estimator
\begin{equation}
E_{ref}(\tau)=E_{ref}(\tau-\Delta\tau)-{{\ln{W(\tau)}-\ln{W(\tau-\Delta\tau)}}\over{\Delta\tau}},
\label{eq:E_ref}
\end{equation}
where $W(\tau)=\sum_k w_k(\tau)$.
Dependence on the size of the time step $\Delta\tau$ has been carefully analyzed in order to
eliminate any significant bias. A resulting optimal value of $\Delta\tau$=50 a.u. is
used. 
After performing a thousand steps to equilibrate the initial ensemble, 
averages are calculated by performing block averaging on blocks consisting of 500 steps, 
and subsequent averaging of these block averages.  

\subsection{Excited state calculations}
\subsubsection{Projection operator imaginary time spectral evolution (POITSE) method}
The POITSE method allows the calculation of excited state energies
from a stochastic solution of the Schr\"odinger equation without imposing
any nodal approximations.
In this scheme, the excited state energies are extracted from the two-sided inverse Laplace
transform of an imaginary time correlation function $\tilde{\kappa}(\tau)$ that
is obtained by a multi-dimensional Monte Carlo integration combined with
zero temperature diffusion Monte Carlo sidewalks.
The basic formulation of the method has been previously discussed \cite{blume97},
and thus we present here only a brief summary of the main ideas, together
with details specific to the present study.

The analysis starts with the spectral function
\begin{equation}
\kappa(\omega) = \sum_n \mid \langle \psi_0 \mid \hat{A} \mid \psi_n \rangle \mid^2
                 \delta(E_0-E_n+\omega),
\label{eq:spect_func}
\end{equation}
where $\{\mid\psi_n\rangle\}$ and $\{E_n\}$ are a complete set of eigenfunctions and
eigenvalues for the Hamiltonian $\hat{H}$, respectively. $\hat{A}$ is a local operator
that projects, at least approximately, $\mid\psi_0\rangle$ into the particular
excited state of interest $\mid\psi_n\rangle$. Performing a two-sided Laplace
transform of eq.~(\ref{eq:spect_func}) results in the imaginary time correlation
function
\begin{eqnarray}
\tilde\kappa(\tau)&=&\int_{-\infty}^{+\infty} \exp{(-\omega\tau)} \kappa(\omega) d\omega
\label{eq:laplace} \\
&=&\langle \psi_0 \mid \hat{A} \exp{[-(\hat{H}-E_0)\tau]} \hat{A}^{\dag} \mid \psi_0 \rangle
\label{eq:corr_func0} \\
&=&\sum_n \mid \langle \psi_0 \mid \hat{A} \mid \psi_n \rangle \mid^2 e^{-(E_n-E_0)\tau}.
\label{eq:corr_func1}
\end{eqnarray}
\noindent
The matrix element of eq.~(\ref{eq:corr_func0}) involves
only the ground state and, therefore,
it can be computed by using a multi-dimensional Monte Carlo (MC) integration over the
distribution $\mid\psi_0\mid^2$ and then performing a DMC sidewalk at each MC point.
In most cases, however, the exact ground state $\mid\psi_0\rangle$ is not known.
For practical purposes, one typically employs a trial function $\mid\Psi_T\rangle$ and
a reference energy $E_{ref}$. These approximate $\mid\psi_0\rangle$ and
$E_0$, respectively, as closely as possible. Use of a reference energy not equal to the exact ground state
energy modifies the decay rate of all terms in eq.~(\ref{eq:corr_func1}) by a constant factor
$E_{ref}-E_0$. It has been shown in ref.~\cite{blume97} that this dependence on the arbitrary
reference energy can be eliminated by introducing the normalization factor
\begin{equation}
\langle \Psi_T \mid \exp{[-(\hat{H}-E_{ref})\tau]}
                          \mid \Psi_T \rangle.
\label{eq:norm_fact}
\end{equation}
Thus, replacing $\mid\psi_0\rangle$, $E_0$ in eq.~(\ref{eq:corr_func0}) with $\mid\Psi_T\rangle$,
$E_{ref}$, respectively, and dividing by the additional normalization factor of
eq.~(\ref{eq:norm_fact}), leads to the desired time correlation function
\begin{equation}
\tilde\kappa(\tau)=\frac{ \langle \Psi_T \mid \hat{A} \exp{[-(\hat{H}-E_{ref})\tau]}
                          \hat{A}^{\dag} \mid \Psi_T \rangle}
                        { \langle \Psi_T \mid \exp{[-(\hat{H}-E_{ref})\tau]}
                          \mid \Psi_T \rangle}.
\label{eq:normcorr_func} \\
\end{equation}
\noindent
From eq.~(\ref{eq:corr_func1}) it is clear that the imaginary time decay of 
$\tilde{\kappa}(\tau)$ contains information
about energy differences $E_n-E_0$.
Therefore calculation of the numerical inverse Laplace
transform of $\tilde{\kappa}(\tau)$ provides
the corresponding
spectral function $\kappa(\omega)$ whose peak positions correspond to
the excitation energies $E_n$.

The first step in the evaluation of $\tilde{\kappa}(\tau)$ is
a variational Monte Carlo walk in which an initial ensemble of
1000 walkers distributed according to $\Psi^2_T$ is generated.
The starting VMC ensemble is then propagated for 8000 steps of $\Delta\tau$=50 a.u. by a DMC
sidewalk, during which eq.~(\ref{eq:normcorr_func}) is sampled.
The trial function is given by eqs.~(\ref{eq:prod_func}-\ref{eq:ocs_func}),
and the VMC and DMC implementations have been described in the 
previous sections.

As well known, the numerical inversion of $\tilde{\kappa}(\tau)$ to obtain
$\kappa(\omega)$ is an
ill-conditioned problem, especially when Monte Carlo noise is non-negligible and/or
when the spectral function $\kappa(\omega)$ contains multiple overlapping peaks of
comparable intensity. Thus a judicious choice of the operator $\hat{A}$ is necessary
to ensure that the time-dependence of $\tilde\kappa(\tau)$ is dominated by only one
or few well-separated energy differences.

In order to compute the rotational excitations of OCS inside $p$-H$_2$ clusters we employ here
projectors $\hat{A}$ proportional to the molecular Wigner functions $D_{mk}^{j}(\alpha,\beta,\gamma)$.
Here, $\alpha,\beta,\gamma$ are the three Euler angles that define the orientation of the 
molecular frame in the arbitrary space-fixed frame, and $j,m,k$ are the OCS rotational quantum numbers.
We focus in particular on $D_{00}^j$, with $j$ ranging from 1 to 5.
These projectors, except for a numerical constant, correspond to the rotational
eigenfunctions of a linear molecule. They are also equivalent to the eigenfunctions
of a symmetric top rotor with $k$=0. 
They are functions only of the second Euler angle and are expressed in terms of
the Legendre polynomials, i.e., $D_{00}^j \propto P_j(\cos\beta)$. 
Therefore, $\hat{A}$ accesses states in which the total angular momentum $J$ is carried
primarily by the OCS molecule, i.e. $J\approx j$.

The inverse Laplace transform of
$\tilde{\kappa}(\tau)$ is performed here using the implementation of the Maximum
Entropy (MaxEnt) method as described in ref.~\cite{bryan90}. 
Our use of this approach 
is identical to that employed in previous POITSE applications \cite{blume97,blume97a,blume99,huang03}.
Since the maximum entropy
analysis requires independent samples of $\tilde{\kappa}(\tau)$, the initial configuration
for each DMC sidewalk is taken from the VMC walk every 200 steps apart, to minimize
correlations between successive sidewalks. In the present study, 8000-10000 independent decays
are required to produce a converged spectral function $\kappa(\omega)$ for all the projectors
defined before.
We have also calculated the excitation energies by fitting the imaginary time correlation function 
with a sum of exponentials as was done in ref.~\cite{moroni03}, i.e.,
\begin{equation}
\tilde{\kappa}(\tau)=\sum_{n} c_{n} e^{-(E_n-E_0)\tau}.
\label{eq:exp_fit}
\end{equation}
Typically, three terms in this sum are sufficient to obtain convergent results.
In eq.~(\ref{eq:exp_fit}), the coefficient $c_n$ is the spectral weight 
of the particular excitation $E_n$ contributing to $\tilde{\kappa}(\tau)$. 
In order to discuss the convergence of our results we also analyse the 
quantity $\epsilon_J(\tau)$ defined as \cite{moroni03}
\begin{equation}
\epsilon_J(\tau)=-{{1}\over{\tau}}\ln{[\tilde{\kappa}^{(J)}(\tau)/c^{(J)}_1]},
\label{eq:epsilon}
\end{equation}
where $c^{(J)}_1$ is the largest spectral weight obtained from the exponential fit of
eq.~(\ref{eq:exp_fit}).

The analogies and the differences
between MaxEnt and the exponential fit are discussed in more detail in section IIIB.

\subsubsection{Clamped coordinate quasiadiabatic diffusion Monte Carlo}
Excited state energies can also be calculated using the clamped coordinate quasiadiabatic
diffusion Monte Carlo (ccQA-DMC) method proposed by Quack and Suhm \cite{quack91}.
We have previously applied this approach to analysis of small OCS($^4$He)$_N$ clusters \cite{paesani01b} 
and the OCS($p$-H$_2$) complex \cite{zillich03}.  

The key idea of this approximate approach is the assumption that the whole cluster follows the rotation
of the dopant molecule. In the present implementation
the instantaneous inertial tensor $I(\tau)$ is computed at each step of the DMC propagation.
Diagonalization of $I(\tau)$
provides an effective centrifugal potential $V_{J,K_a,K_c}({\bf R})$,
where $J$ is the total angular momentum quantum number, and $K_a, K_c$ are the pseudo-quantum numbers
of an asymmetric rotor \cite{kroto}. 
Adding $V_{J,K_a,K_c}({\bf R})$ to $V({\bf R})$ of eq.~(\ref{eq:potential}) leads to a modified 
Hamiltonian operator
\begin{equation}
\hat{H}_{J,K_a,K_c}({\bf R})=\hat{H}({\bf R})+V_{J,K_a,K_c}({\bf R}).
\label{qa}
\end{equation}
The RB-DMC method can then be used to solve the imaginary time 
Schr\"odinger equation for $\hat{H}_{J,K_a,K_c}({\bf R})$,
given a fixed set of $J$, $K_a$ and $K_c$.
Furthermore, because $\hat{H}({\bf R})$ and $\hat{H}_{J,K_a,K_c}({\bf R})$ 
differ only by the centrifugal
potential term, a correlated sampling scheme \cite{wells85} can be used to calculate  
$E_0$ together with all excitation energies $E_{J,K_a,K_c}$ in a single DMC run. 
This significantly reduces the statistical errors and also reduces the 
computational cost.

Since the instantaneous inertial tensor is a coordinate operator, 
this is implicitly determined by the mixed distribution $\Psi_T \Phi_0$ (see eq.(\ref{eq:mix})).  
Consequently the ccQA-DMC excited states energies may be indirectly subject 
to some trial function bias.  
We discuss this and other possible sources of inaccuracy in the ccQA-DMC 
approach in Section~\ref{subsec:ccQA_results}.

\section{Results}
\subsection{Ground state}
Both the VMC and DMC ground state energies for the OCS($p$-H$_2$)$_N$ clusters
are listed in Table~\ref{tab:energy}. We also report there the energy per 
$p$-H$_2$ molecule derived from the DMC results (column 4). The total energy decreases as 
a function of $N$, indicating that the cluster is stabilized by adding $p$-H$_2$ molecules.
One can also see that up to $N$=5 the difference between DMC and VMC energies
is less than 5$\%$. It subsequently increases for the largest sizes, illustrating
the difficulty of designing accurate trial functions for clusters with $N\geq 6$.
The fact that $N$=5 corresponds to some "magic" number is supported by inspection
of the energy per single $p$-H$_2$ molecule. This quantity first decreases with $N$, 
reaching a minimum value at $N$=5 and then rapidly increases.
A similar behavior was seen in the energetics for
OCS($^4$He)$_N$ clusters \cite{paesani01}.

The difference between clusters with $N\leq 5$ and those with $N>5$ 
becomes
more evident if one considers the para-hydrogen chemical potential $\mu_N=E_0(N)-E_0(N-1)$.
This is reported in Fig.~\ref{fig:chem_pot}. After decreasing up to 
$N$=4, $\mu_N$ remains approximately constant at $N$=4-5. It then markedly        
increases when $N$=6. For $N$=7-8 $\mu_N$ is again constant at a value slightly below the value 
found for $N$=6. This indicates that the OCS($p$-H$_2$)$_6$ cluster is relatively less stable than 
the others. 

The above energetic analysis suggests
that some structural change might be occuring
between $N$=5 and $N$=6. This hypothesis is confirmed on inspection of Fig.~\ref{fig:2D_rho} 
where we show 2-dimensional contour plots of the number density $\rho(r,z)$ defined in 
eq.~(\ref{eq:2D_rho}). 
Up to $N$=5 the ground state structure of the clusters corresponds to
a single ring of para-hydrogen molecules around the OCS axis. 
For these small sizes the $p$-H$_2$ molecules are all located in
the global minimum region of the OCS-($p$-H$_2$) interaction (see Figs.~2a and 2b).
From $N$=6 onwards, a second ring appears in the region
close to the oxygen side. Note that this ring is not
located in either of the two collinear local minima.
Integration of $\rho(r,z)$ for the $N$=8 cluster shows
that five $p$-H$_2$ molecules are still located in the first ring, while the other three are
found in the second one.
Note that while the first (lowest) local minimum of the OCS-($p$-H$_2$) interaction
potential is located at the sulfur end ($\vartheta$=0$^\circ$) and is about 20 cm$^{-1}$ deeper than the second one
at the oxygen end (see Fig.~2b), after the global minimum region
is filled up with five $p$-H$_2$
molecules, no $p$-H$_2$ molecule is found at $\vartheta\approx 0^\circ$.
This can be easily explained by considering the additional effect of the ($p$-H$_2$)-($p$H$_2$) 
interaction. This provides an attractive potential that pulls the
added $p$-H$_2$ molecules closer to the first ring. Because of the saddle point at
$\vartheta$=53.5$^\circ$ the region close to the oxygen side ($\vartheta$=0$^\circ$) is preferred 
over the sulphur side ($\vartheta$=180$^\circ$).

We also note that in the density plots, the peak corresponding to the first ring is separated 
from that of the second ring by a region where the density is 
essentially zero. 
This indicates that after the first ring is completed there is
no delocalization of the other $p$-H$_2$ molecules along the OCS axis.
This is in contrast to what
was previously found for $^4$He in the analogous OCS($^4$He)$_N$ clusters, 
where extensive helium delocalization is evident
\cite{paesani01,paesani01b}.

Although significant information about the cluster structures can be extracted from 
the number density $\rho(r,z)$,
this does not provide any information concerning the localization of
the $p$-H$_2$ molecules in planes perpendicular to the OCS axis. 
In the left panels of Fig.~\ref{fig:corr_fun} we therefore
show the pair angular distribution function defined in eq.~(\ref{eq:pair}), for clusters 
with $N$=2-6. 
In the right panels of the same figure we also report for comparison the corresponding
results obtained for the same size OCS($^4$He)$_N$ clusters. In this case
we have used the He-OCS interaction potential of ref.\cite{higgins99} and the 
He-He potential proposed by Aziz and co-workers \cite{aziz87}.
$P_{12}(\phi)$ describes here the pair distribution of para-hydrogen (helium)
in the ring located in the global minimum region of the OCS-($p$-H$_2$) 
(OCS-He) interaction.

At first sight it is evident that the OCS($p$-H$_2$)$_N$ clusters show a stronger localization
of the $p$-H$_2$ molecules in this ring.
It is also important to note that up to $N$=5 there is a zero probability 
to find two $p$-H$_2$ molecules 
closer than their repulsive hard-core diameter ($\sim3$~\AA). This evidence together with the
fact that this ring is finite and effectively one-dimensional implies
that only cyclic permutations among $p$-H$_2$ molecules are possible.
For larger $N$, instead, the probability to occupy sites outside the global potential well
is finite (see Fig.~\ref{fig:2D_rho}), and this results in a
finite value of $P_{12}(\phi)$ at $\phi$=0$^\circ$.

The analysis of the pair angular distribution also provides some insight into
the relative rigidity of the first ring as a function of the number
of $p$-H$_2$ molecules.
For $N$=2, $P_{12}(\phi)$ shows two distinct peaks and thus the structure of the 
cluster can be reasonably described as a "rigid" $p$-H$_2$ dimer that is located
perpendicular to the OCS axis.
When $N$=3-4 the ring is instead "floppy", as evidenced by the
broad central band in $P_{12}(\phi)$. In particular, we note that for $N$=4
there are not three distinct peaks corresponding to three $p$-H$_2$
molecules located at well-defined angle $\phi$ with the fourth at $\phi$=0$^\circ$.
This means that the four $p$-H$_2$ molecules are not arranged in a square
configuration around the OCS, i.e., the ground state structure of OCS($p$-H$_2$)$_4$
is not exactly symmetric. 

The most interesting aspect, however,
is the structural transition that occurs when $N$=5.
For this cluster size, in fact, $P_{12}(\phi)$ looks entirely different.
This now shows
four sharp peaks,
corresponding to four well-localized $p$-H$_2$ with the reference $p$-H$_2$ 
at $\phi$=0$^\circ$. 
The OCS($p$-H$_2$)$_5$ structure can thus be rationalized as a "rigid"
ring of para-hydrogen molecules around the OCS axis. In contrast,
the OCS($^4$He)$_N$ clusters appear to be delocalized in
the $\phi$ degree of freedom and, consequently, very floppy for all the sizes
shown in Fig.~\ref{fig:corr_fun}.

\subsection{Excited states}
\subsubsection{POITSE results}
In Section III.A  we have seen that the first few para-hydrogen molecules form
a ring around the OCS axis. However, as discussed there, for $N<5$ 
the ring is not complete and
the $p$-H$_2$ are not symmetrically arranged (see Fig.~\ref{fig:corr_fun}).
This implies that the molecular axis is not the axis of symmetry of the whole 
cluster which, as a result, has an asymmetric structure.

As well known, the rotational eigenfunctions for an asymmetric rotor
are given by a linear combination of the Wigner functions \cite{zare}.            
In POITSE calculations, however, we use projection operators that are proportional only
to one of these functions, namely $D_{00}^j(\alpha,\beta,\gamma)$. Furthermore, 
the three Euler angles are defined with respect to a frame fixed on the OCS molecule
that, as noted above, does not coincide with the principal axis frame
of the whole cluster. As a consequence of this, the excited state wavefunction obtained
after the action of $D_{00}^j(\alpha,\beta,\gamma)$ on the ground state wavefunction, is not
an eigenstate of the Hamiltonian. 
Therefore, it can have a non zero overlap with 
excited states different from those we are considering here. In a symmetric top
limit, these states correspond to states with $J$=$j$ and $K$=$k$=0, 
where $J$ is the total angular momentum quantum number, 
and $K$, $k$ are the quantum numbers for the total angular momentum projection along 
the space-fixed and the body-fixed $z$ axis, respectively. 

That this is indeed the case we are facing here becomes clear if one looks
carefully at Fig.~\ref{fig:spec1_4}.
In the right panels, we show the spectral
function obtained from the Maximun Entropy algorithm 
for $N$=1-4 and $J$=1-3. In the left panels we show
the corresponding quantity $\epsilon_J(\tau)$ (solid lines) of eq.~(\ref{eq:epsilon}), 
and the excited state energies (dotted lines)
obtained from the exponential fits, eq.~(\ref{eq:exp_fit}).
 
It is evident that for $J$=1 the quantity $\epsilon_J(\tau)$ is constant in time
for all the four clusters, indicating that the projection operator $D_{00}^{1}$ 
accesses primarily one eigenstate of the Hamiltonian. 
However, for $J$=2 and $J$=3,  
$\epsilon_J(\tau)$ first increases, reaches a maximum value corresponding
to the excitation with the largest spectral weight, and then 
decreases, as a function of $\tau$. Furthermore,
the larger is $J$, the more pronounced is this last decrease. 
This is the consequence of the fact that, 
because of the asymmetric structure of these smallest clusters,
the projection operators $D_{00}^{j}$ can access different rotational excited states.
Thus the corresponding
imaginary time correlation function contains more than one eigenergy
and $\epsilon(\tau)$ varies as these eigenenergies are accessed. 
It is also important to note that when $N$=4, the OCS axis 
more closely approximates the axis of symmetry for the whole cluster, 
and hence $\epsilon(\tau)$ becomes constant in time also for the higher 
values of $J$.

From these considerations, it follows that the inversion of the correlation
function $\tilde\kappa(\tau)$ is technically more difficult for 
the highest values of $J$ and for $N\leq 3$.
In fact, while for $J$=1 the result of inversion is independent of the time $\tau$, 
for $J$=2 and $J$=3
MaxEnt provides a stable spectral function with a narrow peak only 
when the inversion range is restricted to the time interval
before $\epsilon(\tau)$ decreases.
The same sensitivity to time interval is found when 
the inversion is perfomed using the exponential fit.
Identical results are obtained from the two inversion procedures.
However, in the case of the exponential fit,
the optimal fit results in a very large $\chi^2$ value 
($\chi^2\approx 5\times 10^5$), indicating that 
caution must be used when dealing with this procedure.

From $N$=5 onwards, we have seen that a rigid ring of five $p$-H$_2$ 
molecules is formed around the OCS axis
and thus the molecular axis can be reasonably identified with 
the axis of symmetry of the whole cluster.
As a consequence of this, the $D_{00}^j$'s become "good" 
projection operators each of which accesses only one eigenstate of the Hamiltonian. 
The inversion of $\tilde\kappa(\tau)$ is thus simpler
for these sizes and we are able to extract the excited state energies for $J$
ranging from 1 to 5.
In Fig.~\ref{fig:spec5_8} we report, as an example, 
$\epsilon_J(\tau)$ and $\kappa(\omega)$ for $N$=5 and $N$=8.
From the use of the Maximum Entropy method we have found that, 
while the peak positions are independent of the
length of the time interval,
narrower peaks can be obtained if only the central part (from $\tau$=1.0$\times$10$^5$ a.u. to
$\tau$=3.0$\times$10$^5$ a.u.) of the
time decay is considered. 
For these cluster sizes the exponential fit also provides identical results as MaxEnt, but
again the associated $\chi^2$ is very large (see above).

This detailed analysis of the imaginary time dependence of $\epsilon(\tau)$ shows that the 
complexes become significantly symmetric for $N\geq 5$.

\subsubsection{ccQA-DMC results}
\label{subsec:ccQA_results}
In Fig.~\ref{fig:qa_en} we show the rotational excitation energies $E_{J,K_a,K_c}$ with $J$=1-3
calculated by use of the ccQA-DMC approach for $N$=1-8. 
Each $E_{J,K_a,K_c}$ exhibits a monotonic decrease as $N$ increases.  
The patterns resemble those of slightly asymmetric prolate rotors.  
From $N$=4 onwards, the energy differences between those excited states
that are degenerate in the limit of a perfect symmetric prolate rotor 
become smaller and smaller. 
This indicates that, as the POITSE results have already shown 
(Figs.~\ref{fig:spec1_4} and~\ref{fig:spec5_8}),  
the clusters become more symmetric. 
Note that $N$=5 is the least prolate structure and most symmetric structure in this size range. 
This can be understood as a consequence of the fact that this cluster possesses a single complete ring.
It is also possible to see from Fig.~\ref{fig:qa_en} that the structural 
transition between $N$=5 and $N$=6 evident in Fig.~\ref{fig:2D_rho} manifests
itself in a change of the slope in $E_{J,K_a,K_c}$. This is particularly evident
for the lowest excited states (bottom panel).

In Fig.~\ref{fig:poitse_qa_en} we compare the exact POITSE with the approximate
ccQA-DMC energy levels for $J$=1-3. 
Because the cluster structures
are those of slightly asymmetric prolate rotors, the excitations extracted from 
the molecular projection operators in POITSE correspond to 
those of states with $J$=$j$, $K_a$=0 and $K_c$=1. 
The two sets of results show generally similar behavior but 
there are nevertheless several significant differences.
For $N$=1 and $N$=2 the POITSE values are systematically higher 
than the corresponding ccQA-DMC values, indicating
that for these sizes the $p$-H$_2$ molecules are not as rigidly bound as
assumed by ccQA-DMC. For $N$=3-4, in contrast, the two sets of results
are identical for $J$=1 but differ increasingly as $J$ increases, 
with the greatest discrepancies being seen at $J$=3.  
Such a pattern of differences is identical to what would be expected 
for excitations lowered by centrifugal distortion below 
the corresponding rigid molecule values.  
This can be rationalized by considering that for these two sizes 
the pair angular distribution function has shown that the structures 
are quite floppy (Fig.~\ref{fig:corr_fun}). 
Consequently,
for higher $J$ values the $p$-H$_2$ density can be increasingly easily distorted. 
The rotational energies indicate that $N$=3 is therefore the floppiest complex.
This effect consistently disappears
for $N$=5 and $N$=6 when a complete ring of five $p$-H$_2$ molecules is formed around the OCS axis.  
At these sizes the POITSE and ccQA-DMC energy levels are essentially coincident, 
implying that the true energy levels (POITSE) correspond now to a rigid complex 
in which all $p$-H$_2$ molecules are rigidly coupled to the OCS rotation (ccQA-DMC). 
We can rationalize these different behaviors for the pairs $N$=3-4 and $N$=5-6 
by a coupling of the ring "breathing" mode to the overall rotation in clusters 
with an incomplete $p$-H$_2$ ring at the global potential minimum (see Fig.~\ref{fig:2D_rho}).  
Such coupling results in a centrifugal distortion of the cluster,
due to expansion of this floppy para-hydrogen ring.
The vibrational "breathing" mode is relatively 
facile for an incomplete ring.  
When this is filled with 5 $p$H$_2$, the ($p$-H$_2$)-($p$-H$_2$) interactions effectively lock 
the para-hydrogen molecules into a more rigid structure (see Fig.~\ref{fig:corr_fun}      
and effectively eliminate centrifugal distortion due to ring "breathing".

For $N$=7-8 quite different behavior is seen.  
Here the POITSE values lie slightly below the ccQA-DMC energies 
for all $J$ values, including $J$=1. 
This does not correspond to the pattern of a centrifugal distortion 
away from a rigid complex. 
The systematically lower POITSE energies for these cluster sizes are somewhat surprising. 
In fact, because of the rigid coupling assumptions implicit in the approach, 
the ccQA-DMC energies are expected to provide a lower limit,  
at least for the lowest value of $J$ where non-rigidity effects are less important.  
Resolution of these differences lies in the implicit dependence 
of ccQA-DMC on a "mixed" ground state density.  
Fig.~\ref{fig:ccQA_densities} compares the $\vartheta$ dependence of the "mixed",
eq.~(\ref{eq:mix}), and 
"pure", eq.~(\ref{eq:pure}), ground state densities for $N$=1-8. 
We see that while for $N$=1-6 these densities are very similar, 
for $N$=7-8 the "mixed" state density overestimates the true density 
in the region between the two rings, at the expense of the second ring density.  
This results in an underestimate of the moments of inertia $I_B$ and $I_C$, 
and hence in an overestimate of the rotational energies obtained from ccQA-DMC.

\subsubsection{Rotational constants}
The rotational excited state energies obtained from the POITSE method 
are fitted using a symmetric top Hamiltonian.
For states with $K$=0, this becomes 
\begin{equation}
E(J)=B_{avg}J(J+1)-DJ^2(J+1)^2.
\label{eq:symm_top}
\end{equation}
Because the ground state cluster structures correspond to those of
slightly asymmetric 
prolate top rotors, at least for the smallest sizes (Figs.~\ref{fig:2D_rho}
and \ref{fig:corr_fun}), the averaged rotational constant $B_{avg}$
refers to $(B+C)/2$.
Both $B_{avg}$ and the distortion constant $D$ are shown as a function of $N$
in Fig.~\ref{fig:poitse_rot}.

The averaged rotational constant $B_{avg}$ decreases as a function of $N$, showing a marked change
in its slope between $N$=5 and $N$=6, exactly when the $p$-H$_2$ molecules begin 
to solvate the oxygen side of the OCS. As we have discussed in the previous section,
at $N$=6, one $p$-H$_2$ molecule occupies a site closer to the OCS axis and more 
distant from the perpendicular rotation axes. 
Consequently the moment of inertia increases providing a larger reduction of $B_{avg}$. 
It is also important to note that a similar magnitude decrease of $B_{avg}$ from $N$=1 to $N$=3 and from
$N$=6 to $N$=8 is found. This suggests that the second ring around the OCS axis has similar
azimuthal distribution features as the first one.

The distortion constant $D$ is $\sim 4\times 10^{-4}$ cm$^{-1}$ for $N$=1, 
appearing somewhat larger than that for the OCS-$^4$He complex \cite{tang02},
although the large error bars make a meaningful comparison impossible.  
$D$ is smaller for $N$=2, nearly zero, providing more evidence that 
the first two $p$-H$_2$ molecules constitute a near "rigid" dimer interacting with the OCS.  
Beyond $N$=2, the distortion constant shows a maximum value at $N$=3 that, as we have already discussed
above (Section~III.B.2),
corresponds to the floppiest structure.  
This is followed by a slightly lower, but still significant, value at $N$=4 and 
a subsequent decrease with $N$, because
of the increasingly rigid packing of $p$-H$_2$ molecules around the OCS. 
The relatively substantial values of $D$ for $N$=3-4 are due to the distortion 
of the ring deriving from coupling to the ring "breathing" mode, 
described in Section~III.B.2 above.  
As noted there, for $N$=5-6 five hydrogen molecules are now locked in the ring, 
which will push this breathing mode higher in energy and reduce its amplitude, 
thereby reducing the effects of its coupling to the cluster rotation.  
The continued decrease of $D$ for larger $N$ ($N$=7-8), is very interesting in this context.  
For $N$=7-8 the second shell contains 2 and 3 para-hydrogen molecules respectively.  
One might therefore expect that for the $N$=8 cluster, 
the "breathing" mode of the second 3-molecule ring would similarly couple 
to the rotation and cause some centrifugal distortion.  
However, interaction of the 3 $p$-H$_2$ molecules in this secondary ring 
with the 5 molecules in the primary ring stabilizes and localizes the secondary ring, 
reducing the effects of any rotational coupling to its breathing mode.  
Consequently, there are no appreciable centrifugal distortion effects due to growth of the second ring.

As a result of the approximations made in the method,
the ccQA-DMC calculations can provide estimates for all three 
rotational constants ($A$, $B$, $C$) within its rigid coupling assumption. 
These are reported in Table~\ref{tab:rot_const} together 
with a comparison of the resulting values of B$_{avg}$ with those obtained from POITSE. 
The ccQA-DMC values confirm the increasingly symmetric prolate
structures as $N$ increases. For $N$=5 the rotational constants $B$ and $C$ differ
by only 0.013 cm$^{-1}$, while for $N\geq 6$ this difference
is less than 0.01 cm$^{-1}$. We interpret these as essentially symmetric top structures.
The differences between the two sets of results for B$_{avg}$ 
are very small for all cluster sizes studied here,
indicating that the rigid coupling approximation made 
in the ccQA-DMC method is generally a good approximation.  
Moreover, the differences seen here are smaller than those found in the 
analogous OCS($^4$He)$_N$ clusters \cite{paesani03,moroni03}, implying that
the complexes with para-hydrogen are generally more rigid than those with helium.
However, the ccQA-DMC approach does rely on several approximations that can cause small inaccuracies.  
We have already noted one problem, namely the implicit dependence on mixed densities 
that affected the ccQA-DMC energies for the $N$=7-8 clusters.  
Another general problem is that the ccQA-DMC method overestimates the degree of 
asymmetry of the clusters, since the averaging of fluctuations can dominate 
the results even for highly symmetric structures.  
Thus for the symmetric OCS($p$-H$_2$)$_5$ cluster, 
for which the ground state structure is clearly symmetric in the azimuthal 
degree of freedom (Fig.~\ref{fig:corr_fun}), ccQA-DMC nevertheless yields $B\neq C$. 
This suggests that some small quantitative inaccuracies might be present for $N\geq 5$.

\section{Relation to Experimental Measurements}
In the following, the present results will be discussed in light of
the existing experimental studies of OCS($p$-H$_2$)$_N$ complexes.

For $N$=1 a direct comparison with the data reported in ref.~\cite{tang02} is possible.
The high resolution spectrum for the OCS($p$-H$_2$) complex was observed and analysed using
the conventional asymmetric rotor Hamiltonian in the $a$-reduced form of Watson \cite{watson}.
From that analysis all the rotational parameters were obtained. Structural properties
as the intermolecular distance $R$ and the Jacobi angle $\vartheta$ were also derived.
As discussed before, the POITSE calculations provide rotational excited energies
corresponding only to states with $J$=$j$, $K_a$=0 and $K_c$=$J$. Therefore,
from this set of calculations it is not possible to extract all the rotational
constants. In Table \ref{tab:rot_nrg1} we report the energies calculated from POITSE
and make a comparison with the corresponding experimental values.
Excellent agreement is found for all the three excited states considered here.
The values for $\langle R \rangle$ and $\langle \vartheta \rangle$ calculated by DMC and reported
in Table \ref{tab:rot_struc} also show a very good agreement
with the corresponding experimentally determined quantities. 
A complete analysis of the ccQA-DMC results for $N$=1 was previously reported \cite{zillich03}.

For $N>1$ no experimental results are available for the pure OCS($p$-H$_2$)$_N$ clusters.
Infrared spectra for these clusters were however observed in pure $^4$He and in mixed 
$^4$He/$^3$He droplets \cite{grebenev00,grebenev01,grebenev02}.
Although a direct comparison with those results is not possible 
because it is difficult to estimate the effect of the surrounding helium droplet, 
we may still make
some qualitative comparisons.
The DMC calculations (ground state and ccQA-DMC) 
provide asymmetric structures for clusters with $N<5$ 
(Figs.~\ref{fig:corr_fun}-\ref{fig:qa_en}, and Table~\ref{tab:rot_const}). 
This is confirmed by analysis of POITSE results for the rotational
excited state energies (Figs.~\ref{fig:spec1_4}-\ref{fig:spec5_8} and
discussion in Section III.B.1). 
The present results are thus in good agreement with the experimental 
observations that allow a fit of the rotational spectrum using a symmetric top Hamiltonian
only for $N\geq 5$ \cite{grebenev01}.  
For $N=5$, the rigidity of the single ring structure seen in the DMC structures and 
in the POITSE energies provides justification for the permutation symmetry arguments 
made to explain the lack of a $Q$-branch in the spectrum at this size \cite{grebenev01}.

It is important to point out that the experimental data in helium droplets suggest that the principal ring
located in the minimum region of the OCS-($p$-H$_2$) interaction should contain 
six $p$-H$_2$ molecules \cite{grebenev02}. This appears to be in contradiction with the present 
results which show the first ring containing only 5 para-hydrogen molecules.  
The difference can be easily explained when the effect of the 
helium density in the droplet is taken into account. 
It has recently been demonstrated with path integral Monte Carlo calculations that in the 
mixed cluster OCS($p$-H$_2$)$_M$($^4$He)$_N$, six $p$-H$_2$ molecules are effectively
arranged in the primary ring \cite{kwon03a}.  A ring of six $p$-H$_2$ molecules is also
found in the pure OCS($p$-H$_2$)$_N$ clusters with $N>$8 \cite{paesani03}. 
The number of para-hydrogen molecules in the primary ring depends 
on the balance of ($p$-H$_2$)-($p$-H$_2$), ($p$-H$_2$)-OCS, and ($p$-H$_2$)-$^4$He interactions.

\section{Summary and conclusions}
The work reported here presents a detailed study of the energetics and 
structures of the OCS($p$-H$_2$)$_N$ clusters with $N$=1-8.
The rigid-body diffusion Monte Carlo method has been employed
to calculate the ground state properties.  
Analysis of the total energy as a function of the cluster size $N$
shows that the clusters are stabilized by adding $p$-H$_2$ molecules.
Analysis of the chemical potential $\mu_N$ indicates that clusters
with $N$=4-5 are relatively the most stable. For these two cluster
sizes the calculation of the $p$-H$_2$ density distribution shows
that a single ring of $p$-H$_2$ is formed around the OCS axis
and is located in the global minimum region of the OCS-($p$H$_2$) interaction. 

Inspection of the $p$-H$_2$ angular distribution inside this ring
provides important informations about its rigidity. For $N$=2
the cluster is effectively constituted by a near "rigid" $p$-H$_2$
dimer bound to the OCS molecule. The clusters with $N$=3-4 appear
to be floppy. When $N$=5 the ring is complete and the $p$-$H_2$
molecules are again "rigidly" arranged around the molecular axis. 
Consequently, the OCS($p$-H$_2$)$_5$ cluster corresponds to a
symmetric structure.

The rotational energies for excited states with $J\approx j$ 
have been also computed as a function of the cluster size $N$ 
by using the POITSE method.
Comparison with results obtained from a rigid coupling approximation,
the ccQA-DMC approach, shows that over this size range the clusters
of OCS with $p$-H$_2$ are to a first approximation fairly "rigid",
showing less deviations from near "rigid" behavior for $N$=6-8 than do 
complexes of OCS with helium~\cite{paesani03,moroni03}.
It has been shown that 
the detailed variations in rotational spectra of the clusters with $N$=1-8 can be 
readily explained in terms of their structural features. 
Thus, for $N\leq 4$ the rotational spectra correspond
to those of slightly asymmetric prolate rotors
with significant centrifugal distortion for
$N$=3-4 that is assigned to coupling of
the rotational motion to a "breathing" mode
vibration of the partially filled para-hydrogen ring.
This coupling is suppressed at $N$=5 when the ring is complete and the ($p$-H$_2$)-($p$-H$_2$)
interactions effectively lock the para-hydrogen molecules into a more rigid structure.
For $N\geq 5$ prolate
symmetric top spectra are therefore obtained with negligible distortion constants. 
Some quantitative inaccuracies in the ccQA-DMC method were seen to arise
when the second ring is growing ($N$=7-8) and when the cluster is highly symmetric ($N$=5).

Comparison with the experimental data for the $N$=1 OCS($p$-H$_2$)
complex shows excellent quantitative agreement for both the energy levels 
and the structural parameters. 
For $N\geq 2$ no experimental data are available for the
pure OCS($p$-H$_2$)$_N$ clusters.  
A qualitative comparison of the main features of the cluster structures 
with conclusions drawn from the experimental results for the 
OCS($p$-H$_2$)$_N$ clusters in the mixed $^4$He/$^3$He
droplets shows good agreement, with the main difference being the 
helium-induced modification of the $N$=6 structure from a 5-membered to 6-membered ring.  
The rigidity of the highly symmetric $N$=5 structure provides strong support 
for the arguments that excited states of this are restricted by nuclear permutation symmetry, 
resulting in a lack of $Q$-branch in the OCS rotational spectrum at low temperatures \cite{grebenev02}.

\begin{acknowledgments}
This work has been supported by the Chemistry Division of the National Science Foundation
(Grant No. CHE-0107541). We thank NPACI for a generous allocation of the computation time 
at the San Diego Supercomputer Center.
\end{acknowledgments}


\newpage
\begin{table}
\caption{Variationally optimized parameters for the trial functions $\xi_T$ and $\chi_T$ of
eqs.~(\ref{eq:prod_func}-\ref{eq:ocs_func}). All values in a.u.
\label{tab:parameters}
}
\begin{tabular}{ccccccccccccc|cc}
$N$ & $p_1$ & $p_2$ & $p_3$ & $p_4$ & $p_5$ & $p_6$ & $p_7$ & $p_8$ & $p_9$ & $p_{10}$ & $p_{11}$ & $p_{12}$ & $p'_1$ & $p'_2$ \\ 
\hline
 1  & -4.35 & 0.793 & 6.12  & -0.22 & -0.16 & 6.38  & 0.958 & -0.188& 0.94  & 1.095    & -0.625   & 0.13     &   -    &   -    \\
 2  & -4.33 & 0.793 & 6.12  & -0.22 & -0.16 & 6.38  & 0.958 & -0.188& 0.94  & 1.095    & -0.625   & 0.13     & 6550   & 0.05   \\
 3  & -4.33 & 0.793 & 6.12  & -0.22 & -0.16 & 6.38  & 0.958 & -0.188& 0.94  & 1.095    & -0.625   & 0.13     & 6550   & 0.05   \\
 4  & -4.33 & 0.793 & 6.12  & -0.22 & -0.16 & 6.38  & 0.958 & -0.188& 0.94  & 1.095    & -0.625   & 0.13     & 6550   & 0.05   \\
 5  & -4.33 & 0.793 & 6.12  & -0.22 & -0.16 & 6.38  & 0.958 & -0.188& 0.94  & 1.095    & -0.625   & 0.13     & 6550   & 0.05   \\
 6  & -4.33 & 0.793 & 6.12  & -0.16 & -0.16 & 6.38  & 0.958 & -0.188& 0.94  & 1.095    & -0.725   & 0.13     & 6550   & 0.05   \\
 7  & -4.33 & 0.783 & 6.12  & -0.05 & -0.16 & 6.38  & 0.958 & -0.188& 0.94  & 1.095    & -0.800   & 0.13     & 6550   & 0.05   \\
 8  & -4.33 & 0.773 & 6.12  &  0.10 & -0.20 & 6.38  & 0.958 & -0.188& 0.94  & 1.095    & -0.800   & 0.13     & 6550   & 0.05   \\
\end{tabular}
\end{table}
\begin{table}
\vspace{2.cm}
\caption{VMC and DMC energies for OCS($p$-H$_2$)$_N$ clusters with $N$=1-8. 
All values in cm$^{-1}$.
\label{tab:energy}
}
\begin{tabular}{cccc}
$N$ &    $E_0^{VMC}$     &    $E_0^{DMC}$     &    $E_0^{DMC}/N$    \\
\hline
 1  &  -74.765$\pm$0.003 &  -74.927$\pm$0.002 &  -74.927$\pm$0.002  \\
 2  & -152.088$\pm$0.007 & -154.206$\pm$0.007 &  -77.103$\pm$0.003  \\
 3  & -231.89$\pm$0.01   & -235.09$\pm$0.01   &  -78.363$\pm$0.003  \\
 4  & -312.98$\pm$0.02   & -319.50$\pm$0.02   &  -79.875$\pm$0.004  \\
 5  & -385.1$\pm$0.03    & -403.57$\pm$0.03   &  -80.715$\pm$0.005  \\
 6  & -427.1$\pm$0.04    & -460.4$\pm$0.05    &  -76.740$\pm$0.008  \\
 7  & -473.1$\pm$0.05    & -521.7$\pm$0.06    &  -74.527$\pm$0.009  \\
 8  & -531.7$\pm$0.05    & -583.5$\pm$0.08    &  -72.94$\pm$0.01    \\
\end{tabular}
\end{table}
\begin{table}
\caption{Rotational constants for OCS($p$-H$_2$)$_N$ clusters with $N$=1-8. 
$B_{avg}$ is defined as $(B+C)/2$.
All values in cm$^{-1}$. The numbers in parentheses indicate the statistical 
error in units of the last digit.
\label{tab:rot_const}
}
\begin{tabular}[t]{c|cccc|c}
&\multicolumn{4}{c|}{ccQA-DMC}&\multicolumn{1}{c}{POITSE}\\
\hline
 $N$  &     A       &     B       &     C      &    B$_{avg}$   &   B$_{avg}$    \\
\hline
  1   & 0.7582(1)   & 0.1902(2)   & 0.1508(2)  &    0.1705(2)   &    0.179(4)    \\
  2   & 0.36756(5)  & 0.1601(1)   & 0.1349(1)  &    0.1475(1)   &    0.152(5)    \\
  3   & 0.23735(3)  & 0.14445(7)  & 0.11680(6) &    0.13062(6)  &    0.136(2)    \\
  4   & 0.17477(3)  & 0.12556(6)  & 0.10699(5) &    0.11627(6)  &    0.120(3)    \\
  5   & 0.13761(2)  & 0.11029(5)  & 0.09797(4) &    0.10413(5)  &    0.108(1)    \\
  6   & 0.12533(3)  & 0.08984(7)  & 0.08052(5) &    0.08518(6)  &    0.088(1)    \\
  7   & 0.11050(3)  & 0.07631(7)  & 0.06806(5) &    0.07218(6)  &    0.068(1)    \\
  8   & 0.09652(3)  & 0.06556(8)  & 0.05923(5) &    0.06240(7)  &    0.0566(9)   \\
\end{tabular}
\end{table}
\begin{table}
\caption{Comparison between POITSE and experimental results [5]
for the OCS($p$-H$_2$) rotational excited state energies
$E_{J,K_a,K_c}$ with $J$=$j$, $K_a$=0 and $K_c$=$J$.
All values in cm$^{-1}$. The numbers in
parentheses indicate the statistical error in units of the last digit.
\label{tab:rot_nrg1}
}
\begin{tabular}[t]{ccccc}
 $J$ & $K_a$ & $K_c$  &     POITSE       &     Exp.       \\
\hline
  1  &   0   &   1    &    0.355(6)      &   0.3534       \\
  2  &   0   &   2    &    1.06(2)       &   1.0574       \\
  3  &   0   &   3    &    2.08(8)       &   2.1068       \\
\end{tabular}
\end{table}
\begin{table}
\caption{Comparison between DMC and experimental results [5]
for the intermolecular distance $\langle R \rangle$ and the Jacobi angle
$\langle \vartheta \rangle$
(see Fig.~\ref{fig:coordinates}). $\langle R \rangle$ is in \AA~ and $\langle \vartheta \rangle$ in degrees.
The numbers in parentheses indicate root mean square deviations.
\label{tab:rot_struc}
}
\begin{tabular}[t]{ccc}
                            &     DMC          &     Exp.       \\
\hline
 $\langle R \rangle$        &    3.794 (0.365) &   3.719        \\
 $\langle\vartheta\rangle$  &    105.7 (9.2)   &   110.8        \\
\end{tabular}
\end{table}

\newpage
\newpage
\begin{figure}[ht]
\centerline{\epsfxsize=15truecm\epsffile{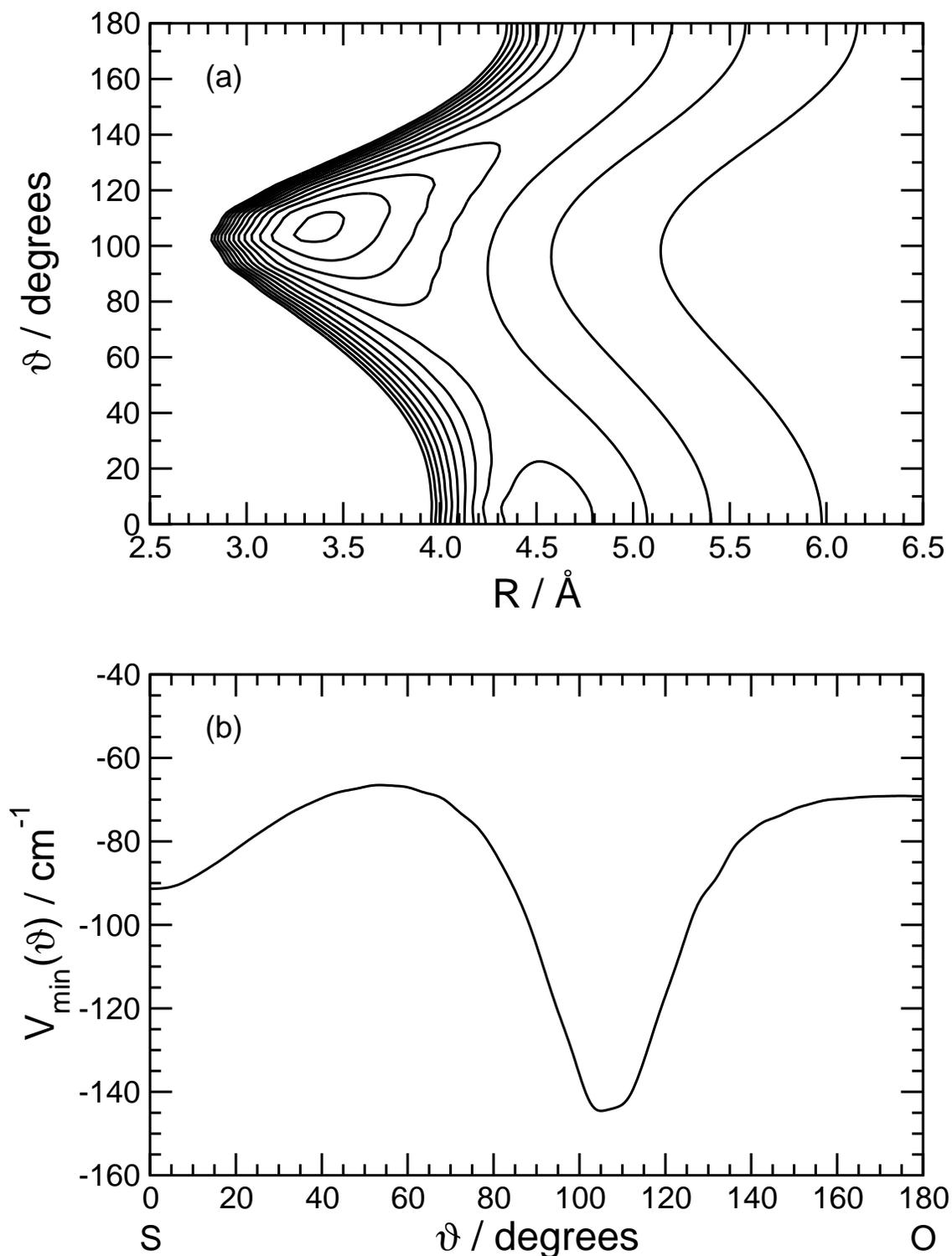}}
\caption{(a) Contour plot of the 2-dimensional OCS-($p$-H$_2$) potential energy surface
obtained by averaging the 4-dimensional potential of Higgins {\em et al.} [15]
over the $p$-H$_2$ angular variables ($\vartheta',\phi'$). 
$R$ and $\vartheta$ are the Jacobi coordinates defined in Fig.~\ref{fig:coordinates}.
The inner isoline around the global minimum at $R$=3.35~\AA and $\vartheta$=105$^\circ$
correponds to -140 cm$^{-1}$. 
The spacing between two consequent isolines is 20 cm$^{-1}$. 
(b) Minimum potential energy path in the angular coordinate $\vartheta$. 
\label{fig:av_pot}
}
\end{figure}
\newpage
\begin{figure}[ht]
\centerline{\epsfxsize=15truecm\epsffile{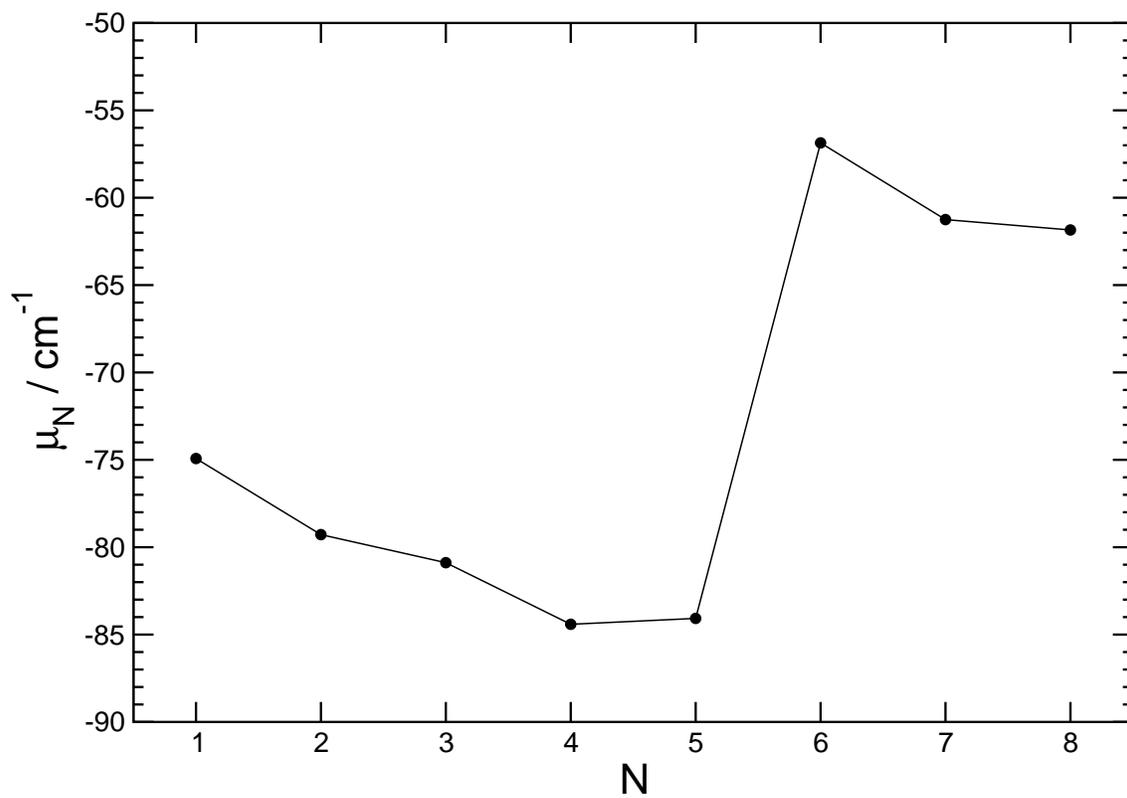}}       
\caption{Chemical potential $\mu_N=E_0(N)-E_0(N-1)$ for OCS($p$-H$_2$)$_N$ clusters 
as a function of $N$. 
All values in cm$^{-1}$.
\label{fig:chem_pot}
}
\end{figure}
\newpage
\begin{figure}[ht]
\centerline{\epsfxsize=18truecm\epsffile{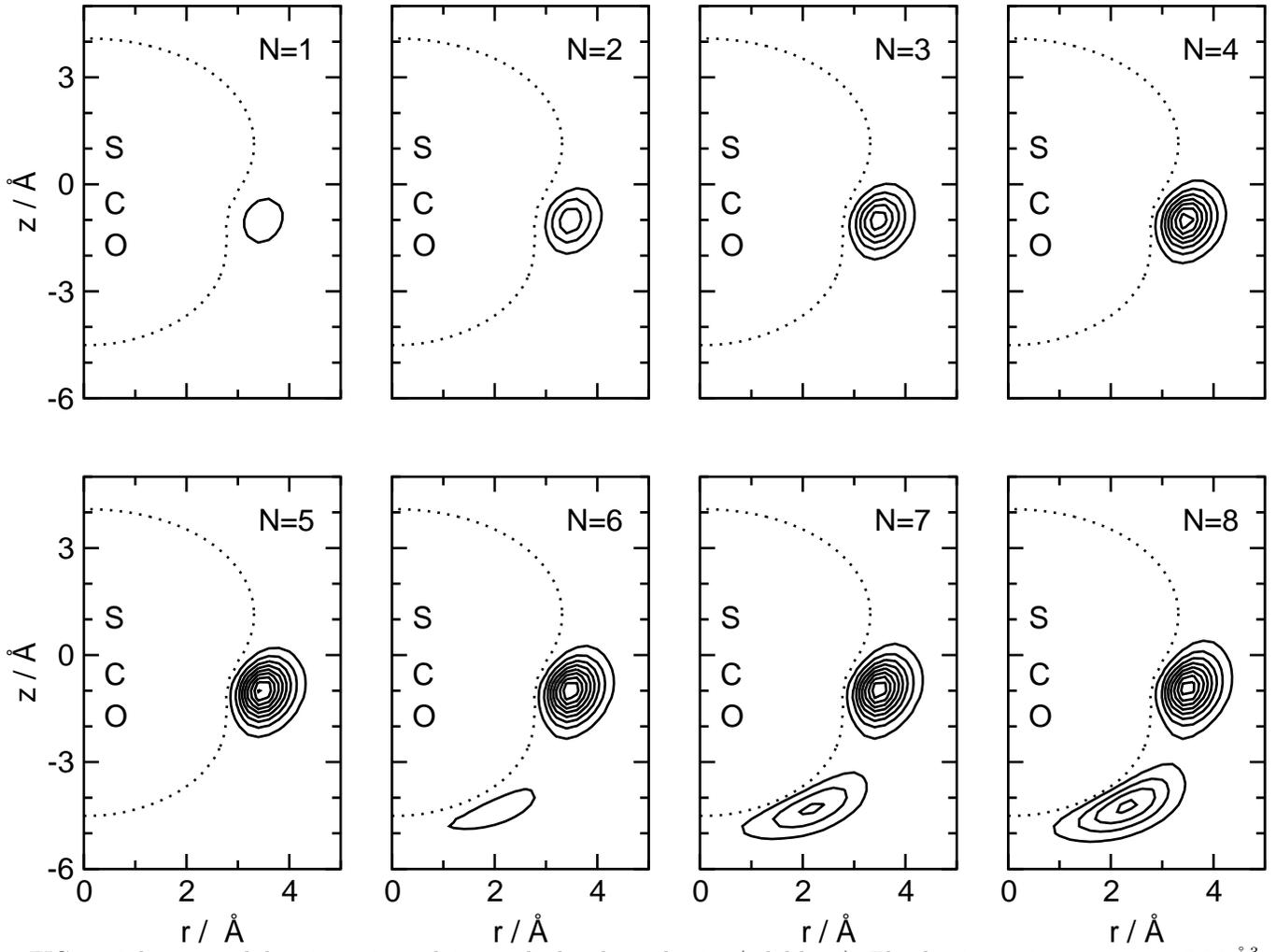}}
\caption{2-dimensional density contour plots in cylindrical coordinates (solid lines).
The density contour spacing is 0.02~\AA$^3$. 
Dotted line: isoline corresponding to V$^{(p-H_2)-OCS}(R,\vartheta)$=0 cm$^{-1}$. 
\label{fig:2D_rho}
}
\end{figure}
\newpage
\begin{figure}[ht]
\centerline{\epsfxsize=18truecm\epsffile{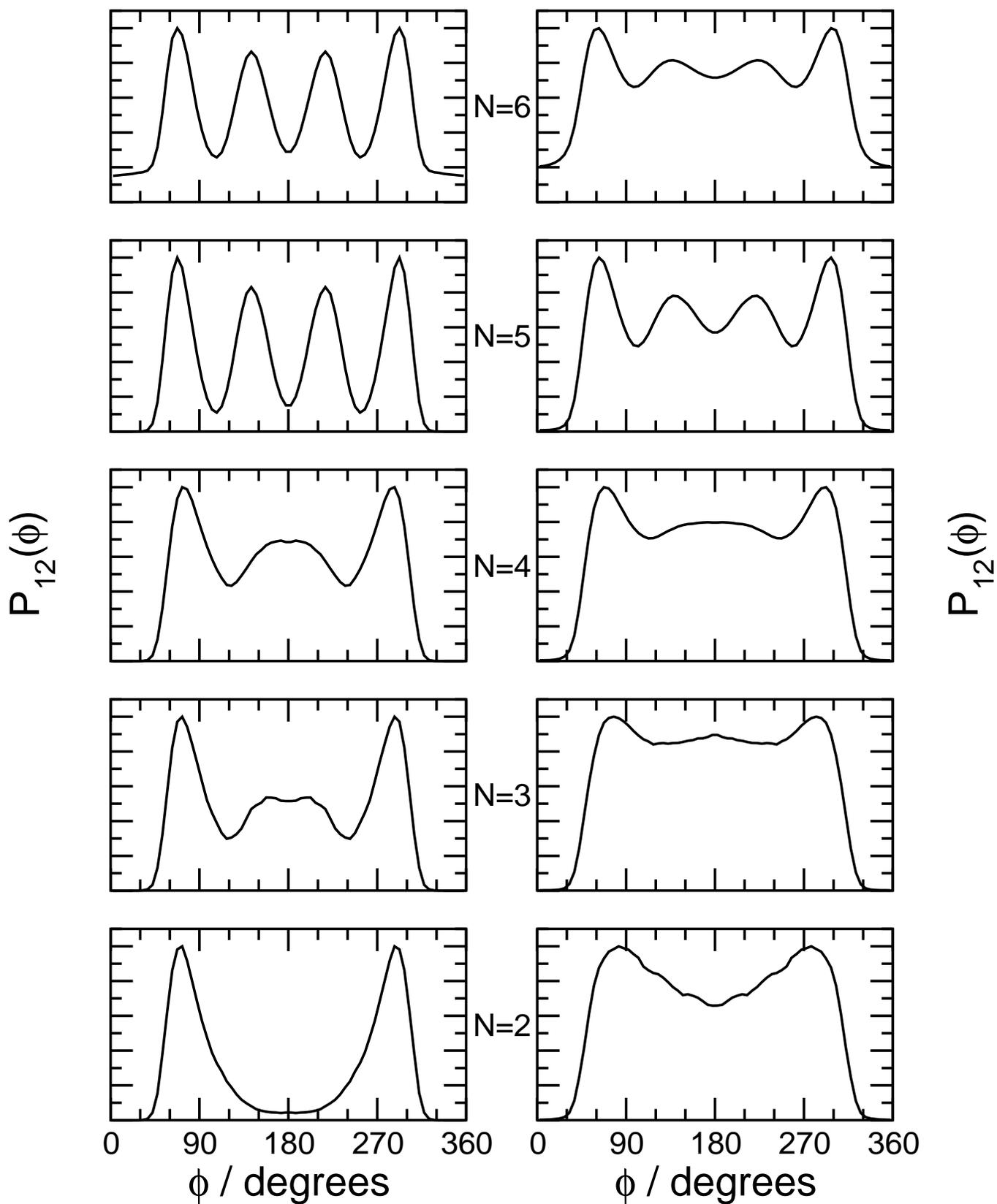}}
\caption{Pair angular distribution function $P_{12}(\phi)$,
eq.~(\ref{eq:pair}), for $N$=2-6.
Left panels: OCS($p$H$_2$)$_N$ clusters, right panels: OCS($^4$He)$_N$ clusters.
\label{fig:corr_fun}
}
\end{figure}
\newpage
\begin{figure}[ht]
\centerline{\epsfysize=21truecm\epsffile{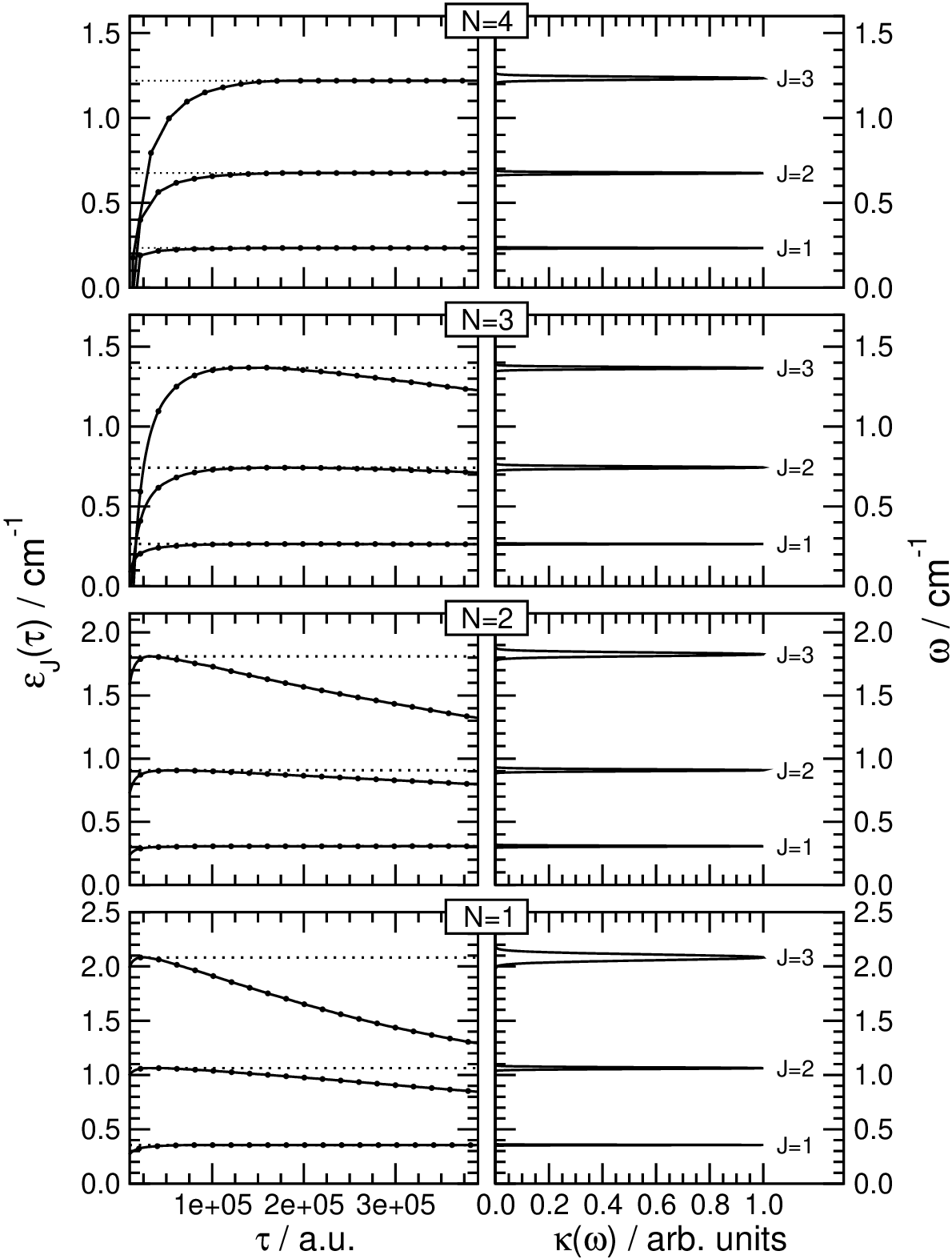}}
\caption{$\epsilon(\tau)$ calculated by means of exponential fit 
eqs.~(\ref{eq:exp_fit},\ref{eq:epsilon}) and spectral function $\kappa(\omega)$ 
eq.~(\ref{eq:spect_func}) calculated using Maximum Entropy method
for $N$=1-4.
\label{fig:spec1_4}
}
\end{figure}
\newpage
\begin{figure}[ht]
\centerline{\epsfysize=21truecm\epsffile{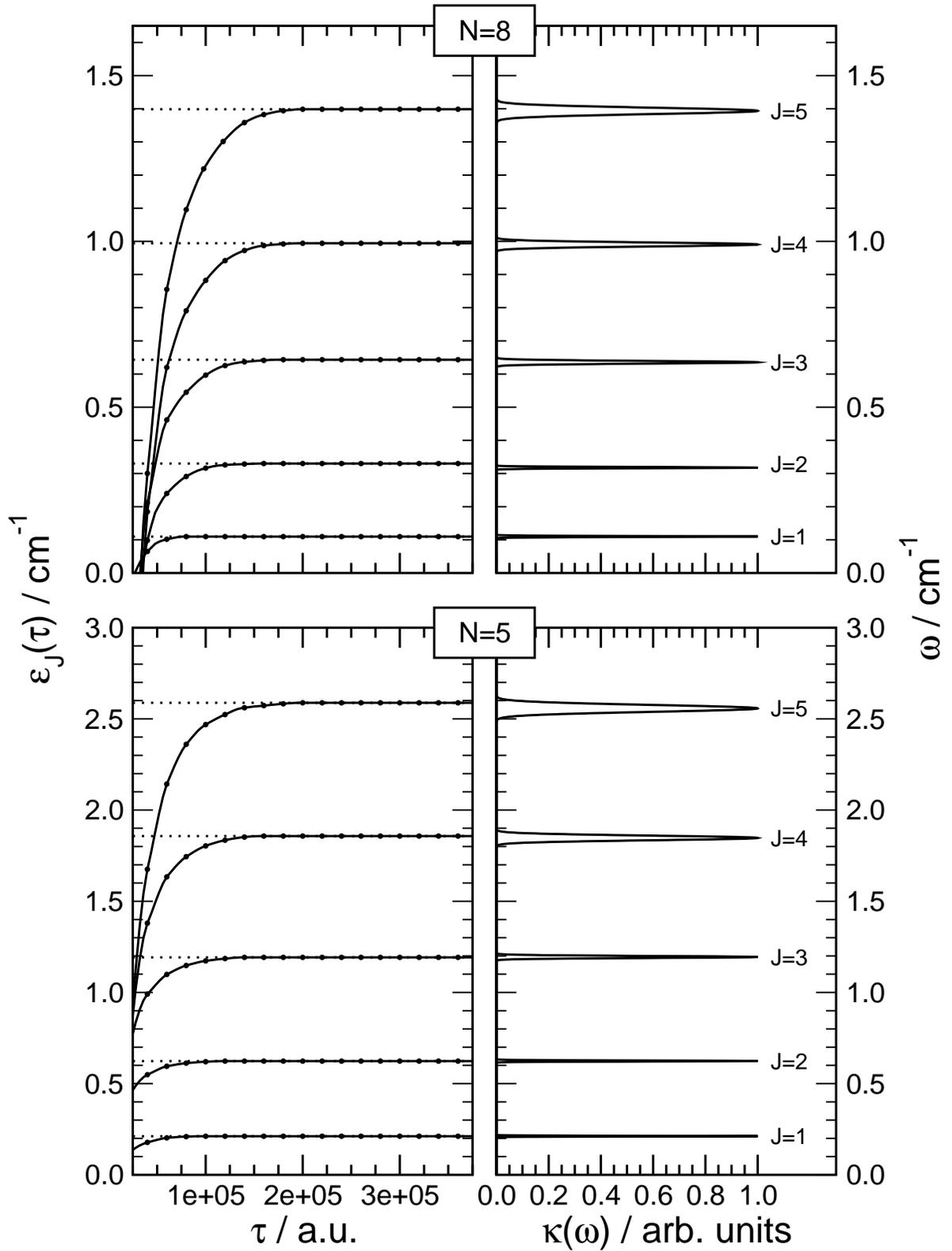}}
\caption{$\epsilon(\tau)$ calculated by means of exponential fit
eqs.~(\ref{eq:exp_fit},\ref{eq:epsilon}) and spectral function $\kappa(\omega)$
eq.~(\ref{eq:spect_func}) calculated using Maximum Entropy method
for $N$=5 and $N$=8.
\label{fig:spec5_8}
}
\end{figure}
\newpage
\begin{figure}[ht]
\centerline{\epsfxsize=15truecm\epsffile{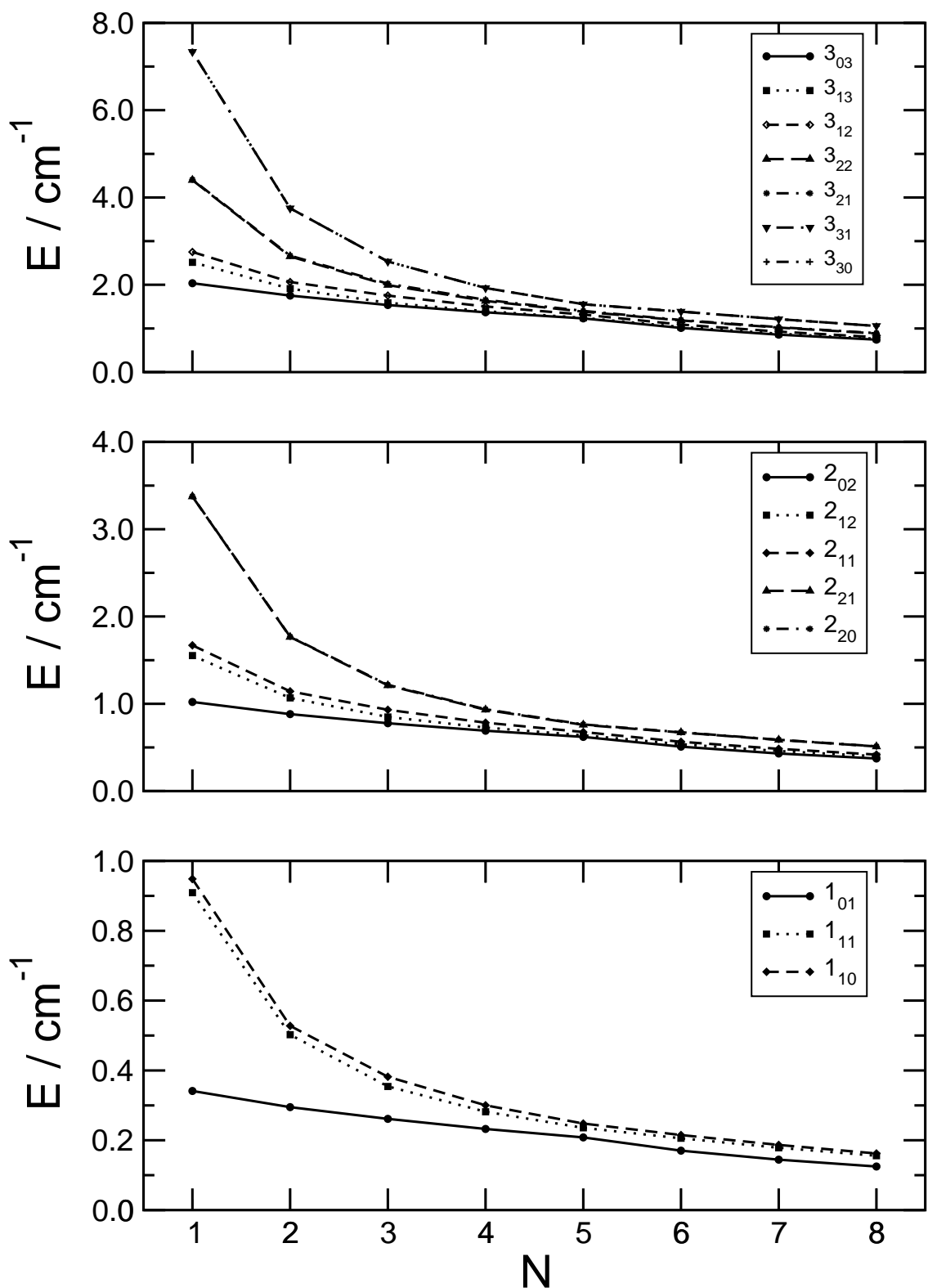}}
\caption{Excited state energies $E_{J,K_a,K_c}$, with $J$=1-3, obtained from ccQA-DMC method. All values 
in cm$^{-1}$.
\label{fig:qa_en}
}
\end{figure}
\newpage
\begin{figure}[ht]
\centerline{\epsfxsize=15truecm\epsffile{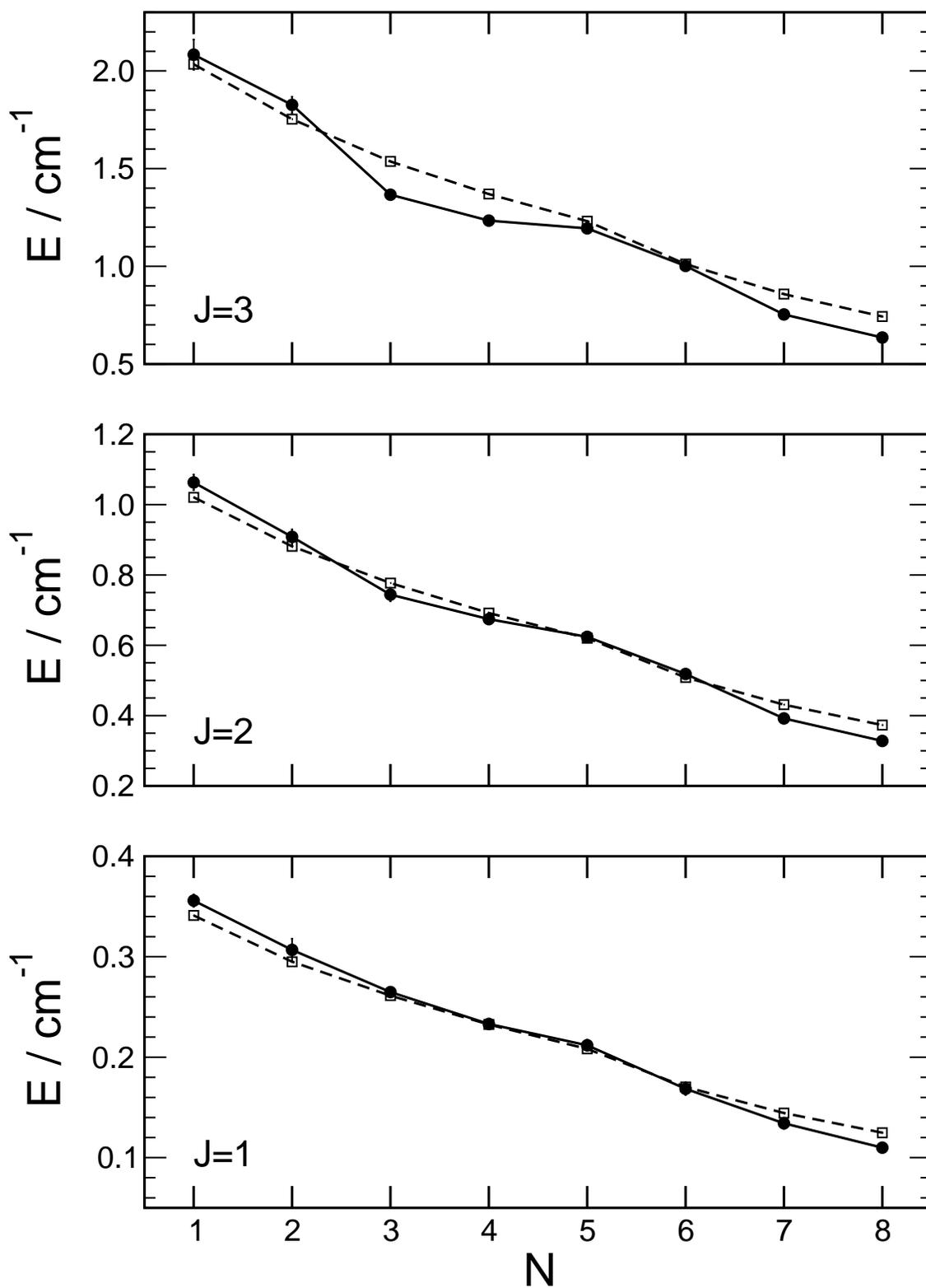}}
\caption{Comparison between POITSE (solid line) and ccQA-DMC (dashed line) energies for 
the rotational excited states of the OCS($p$-H$_2$)$_N$ clusters
with $J$=$j$, $K_a$=0 and $K_c$=$J$. 
\label{fig:poitse_qa_en}
}
\end{figure}
\newpage
\begin{figure}[ht]
\centerline{\epsfxsize=15truecm\epsffile{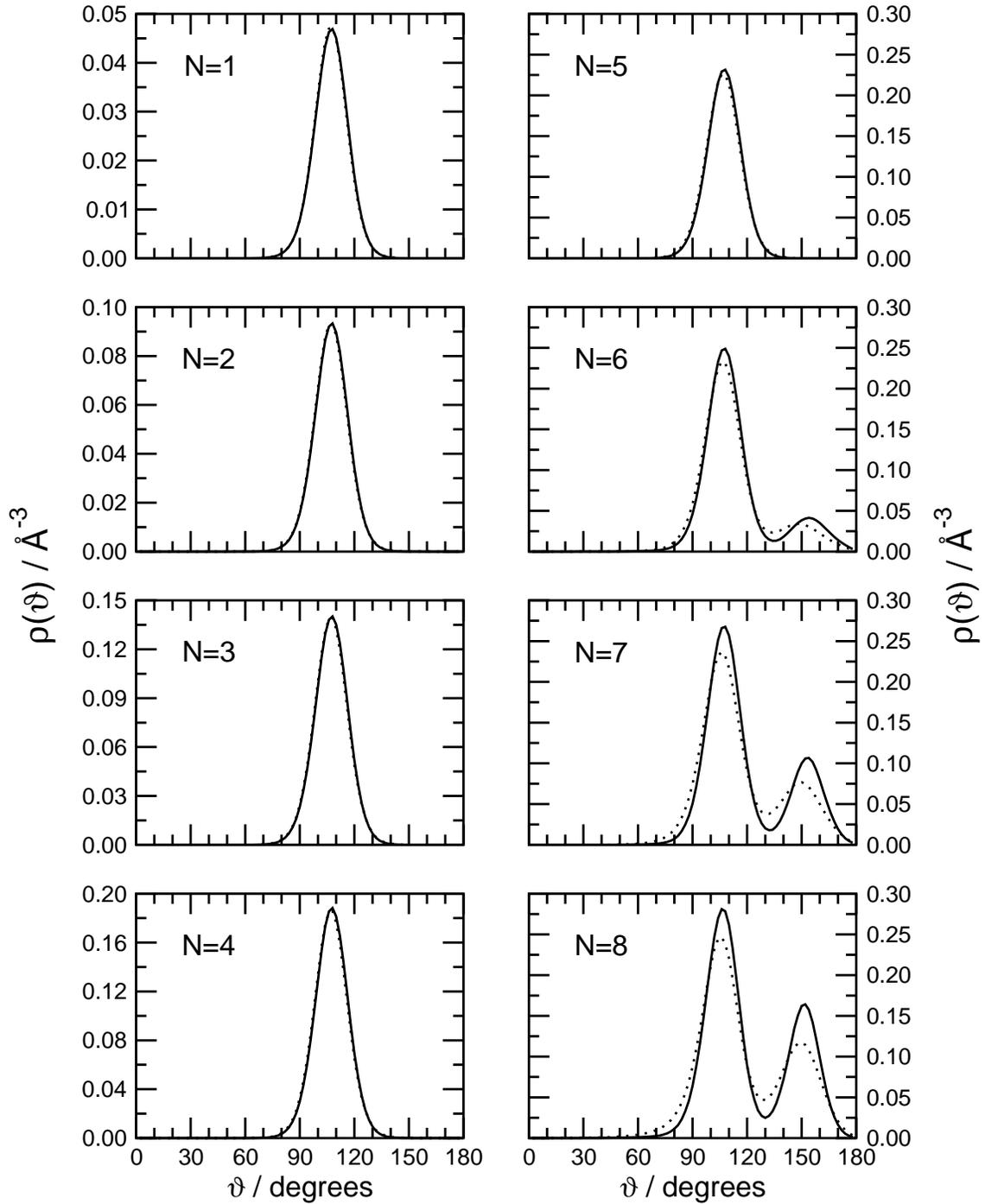}}
\caption{Angular profiles of the $p$-H$_2$ densities for the OCS($p$-H$_2$)$_N$ clusters with $N$=1-8.
Dotted lines: "mixed" densities from eq.~(\ref{eq:mix}), solid lines: "pure" densities from
eq.~(\ref{eq:pure}).
For $N$=1-5 these densities are essentially coincident.
\label{fig:ccQA_densities}
}
\end{figure}
\newpage
\begin{figure}[ht]
\centerline{\epsfxsize=15truecm\epsffile{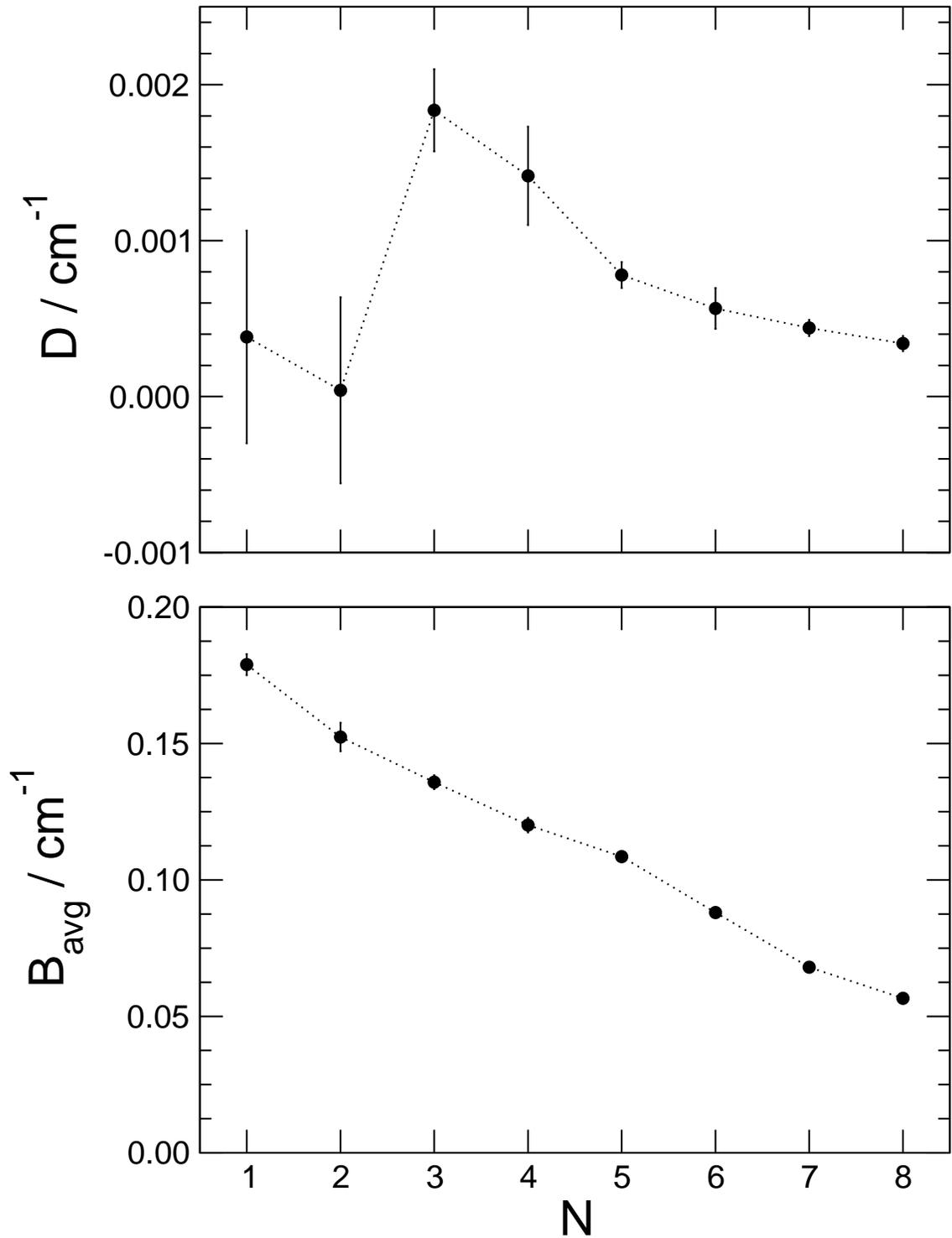}}
\caption{Rotational constant $B_{avg}$ (bottom panel) and distortion constant
$D$ (top panel) obtained from POITSE calculations for OCS($p$-H$_2$)$_N$ clusters
with $N$=1-8.
\label{fig:poitse_rot}
}
\end{figure}

\end{document}